\documentclass[preprint,aps,showpacs,nofootinbib]{revtex4}

\usepackage{epsfig,amssymb,amsmath}

\newcommand{\beq}{\begin{equation}}
\newcommand{\eeq}{\end{equation}}
\newcommand{\bea}{\begin{eqnarray}}
\newcommand{\eea}{\end{eqnarray}}
\newcommand{\bear}{\begin{array}}
\newcommand {\eear}{\end{array}}
\newcommand{\bef}{\begin{figure}}
\newcommand {\eef}{\end{figure}}
\newcommand{\bec}{\begin{center}}
\newcommand {\eec}{\end{center}}
\newcommand{\non}{\nonumber}

\def\EQ#1{Eq.~(\ref{#1})}

\def\REF#1{(\ref{#1})}
\def\GEV#1{10^{#1}{\rm\,GeV}}

\def\oten#1{ {\mathcal O}(10^{#1})}

\begin{document}
\draft
\tighten
\preprint{TU-905}
\preprint{IPMU12-0071}
\title{\large \bf
Anarchy and Leptogenesis}
\author{
    Kwang Sik Jeong$^{(a)}$\footnote{email: ksjeong@tuhep.phys.tohoku.ac.jp}
    and
    Fuminobu Takahashi$^{(a),(b)}$\footnote{email: fumi@tuhep.phys.tohoku.ac.jp}}
\affiliation{
   $^{(a)}$ Department of Physics, Tohoku University, Sendai 980-8578, Japan\\
   $^{(b)}$  Kavli Institute for the Physics and Mathematics of the
 Universe, The University of Tokyo, \\ 5-1-5 Kashiwanoha,
 Kashiwa, Chiba 277-8582, Japan
    }

\vspace{2cm}

\begin{abstract}
We study if leptogenesis works successfully together with the
neutrino mass anarchy hypothesis.
We find that  the predicted neutrino mass spectrum is sensitive to the
reheating temperature or the inflaton mass, while
the distributions of the neutrino mixing angles and CP violation phases
remain intact as determined by  the invariant Haar measure of U(3). In the case of thermal leptogenesis, the light neutrino mass distribution
agrees well with the observations if the reheating temperature is
${\cal O}(10^{9-11})$\,GeV.  The mass spectrum of the right-handed
neutrinos and the neutrino Yukawa matrix exhibit a certain pattern, as
a result of the competition between random matrices with elements of
order unity and the wash-out effect.  Non-thermal leptogenesis is
consistent with observation only if the inflaton mass is larger than
or comparable to the typical right-handed neutrino mass scale.
Cosmological implications are discussed in connection with the
125\,GeV Higgs boson mass.
\end{abstract}

\maketitle


\section{Introduction}
The origin of flavor of quarks and leptons in the standard model (SM)
remains one of the great mysteries in the particle physics: why are
there three generations of quarks and leptons, and why hierarchical
structure in the Yukawa couplings?  There have been proposed a variety
of  models, some of which rely on a hypothetical flavor
symmetry.  However, none of them are decisive as yet. This is partly
because the mass spectrum of elementary particles shows no definite
pattern, unlike the periodic table of elements, and so, if we are to
understand the flavor structure based on the fundamental symmetry
principle, either small symmetry breaking or unknown coupling
constants must be introduced ad hoc, allowing a great variety of
flavor symmetries and charge assignments.

While symmetry has been a useful and attractive guiding principle in
physics, it is not necessarily applicable to all observables. For
instance, the observed vanishingly small cosmological constant may be
interpreted to suggest the existence of some profound fundamental
symmetry setting the cosmological constant (almost) zero.
Alternatively, it may simply be that the cosmological constant is an
environmental parameter adjusted by the anthropic
principle~\cite{Weinberg:1987dv}. Similarly, some of the observables
in our Universe may be strongly affected by the anthropic conditions,
and if so, it is hopeless to try to understand their values from the
symmetry principle. Indeed, many parameters in the SM seem to be
adjusted so that the existence of life is possible. Therefore, the
apparent pattern of the quark and charged lepton mass spectrum may be
just a consequence of the environmental selection, and may not reflect
any fundamental symmetry.\footnote{That said, it is difficult to
  experimentally confirm such anthropic argument. In the rest of this
  paper therefore we do not attempt to interpret the structure of
  Yukawa couplings for quarks and charge leptons. We will come back to
  the flavor symmetry later.}

On the other hand, the situation is different in the neutrino
sector. The neutrinos are massless in the SM, but the neutrino
oscillation experiments have revealed that neutrinos have a tiny but
non-zero mass~\cite{Strumia:2006db}.  Its typical mass is constrained
to be below $0.2$~eV~\cite{Komatsu:2010fb}, and its cosmic mass density
is much smaller than the observed dark matter density. The
tiny neutrino mass can be beautifully explained by the celebrated
see-saw mechanism~\cite{seesaw}; the smallness of the neutrino masses
is related to the ratio of the weak scale and the heavy right-handed neutrino mass
$M_0 \approx \GEV{15}$  close to the GUT scale.  With such tiny mass and cosmic energy density, therefore, the
neutrino mass and mixing may be irrelevant to the existence of life, and so, it may
possess information on its original distribution in the landscape.

Let us briefly summarize the current status of the neutrino parameters.
The three neutrino mixing angles are given by~\cite{Tortola:2012te}:
\bea
\sin^2 \theta_{12} &=& 0.320^{+0.015}_{-0.017},\non \\
\sin^2 \theta_{23} &=& 0.49^{+0.08}_{-0.05} ~(0.53^{+0.05}_{-0.07}),\non \\
\sin^2 \theta_{13} &=& 0.026^{+0.003}_{-0.004} ~(0.027^{+0.003}_{-0.004}),
\label{mixings}
\eea
where the normal (inverted) hierarchy is assumed.
We note that two of them are large, but even the smallest one, $\theta_{13}$,  is not extremely small.
The mass squared differences are~\cite{Tortola:2012te}
\bea
\Delta m_{21}^2 &=& (7.62\pm0.19) \times 10^{-5}\,{\rm eV^2},\non \\
\Delta m_{31}^2 &=& 2.53^{+0.08}_{-0.10}~ (-2.40^{+0.10}_{-0.07}) \times 10^{-3}\,{\rm eV^2}.
\label{mass2d_obs}
\eea
The ratio of the mass squared differences is $\Delta m_{21}^2/|\Delta m_{31}^2| \approx 0.03$, which
is much milder compared to that for the quarks or charged
leptons~\cite{Ibarra:2011gn}. Intriguingly, those neutrino parameters are consistent with
the {\it neutrino mass anarchy hypothesis} proposed in Ref.~\cite{Hall:1999sn},
which has been further studied in
Refs.~\cite{Haba:2000be,deGouvea:2003xe,deGouvea:2012ac}.  In
particular, the recent discovery of non-zero $\theta_{13}$ by Daya-Bay
experiment~\cite{An:2012eh} has made the idea very
attractive~\cite{deGouvea:2012ac}.\footnote{ A hint for non-zero
  $\theta_{13}$ was reported by T2K~\cite{Abe:2011sj},
  MINOS~\cite{Adamson:2011qu} and Double-Chooz~\cite{Abe:2011fz}
  experiments. Recently, RENO experiment also observed the non-zero
  $\theta_{13}$~\cite{Ahn:2012nd}.} As we shall briefly review in the
next section, the neutrino mass anarchy hypothesis  is based on two assumptions:
(i) there is no quantum number to distinguish flavors in the neutrino
sector, and therefore the couplings are structureless in the flavor
space; (ii) the couplings and mass matrix obey basis-independent
random distribution. In particular, it was shown in
Ref.~\cite{Haba:2000be} that the mixing angle distribution obeys the
invariant Haar measure of U(3). Also, using the linear measure of the
eigenvalues of the random matrices, the observed neutrino mass squared
difference can be naturally explained in the see-saw mechanism.  Thus,
both neutrino mass anarchy and the see-saw mechanism are arguably the
most attractive framework for understanding the observed neutrino
parameters.

The origin of matter remains a puzzle in cosmology and particle
physics. Since any pre-existing baryon asymmetry would be
exponentially diluted by the subsequent inflationary expansion, it is
necessary to generate the baryon asymmetry after inflation. One
plausible explanation is the baryogenesis through
leptogenesis~\cite{Fukugita:1986hr}: the lepton asymmetry generated by
the out-of-equilibrium decay of the right-handed neutrinos is
transferred to the baryon asymmetry via the sphaleron process.
However, if the leptogenesis is responsible for the observed matter
asymmetry, it might select a certain subset of the neutrino
parameters, and as a result, the original distribution in the
landscape may be significantly distorted, spoiling the success of the
neutrino mass anarchy.

In this paper we study if the leptogenesis works successfully together
with the neutrino mass anarchy hypothesis.  The result is
two-fold. First, the mixing angles and the CP violation phases (one
Dirac and two Majorana) in the low energy are subject to the invariant
Haar measure of U(3), and they are not modified by requiring the
successful leptogenesis. In a sense, the mixing angles as well as the
CP violation phases are orthogonal to leptogenesis. Second, the
neutrino mass eigenvalues are generically affected by leptogenesis.
We find however that thermal leptogenesis is possible without
significant modification of the predictions of the original neutrino
mass anarchy, if the reheating temperature is
$\oten{9-11}$\,GeV and if the typical right-handed neutrino mass scale is of $\oten{15}$\,GeV.
This is the result of the competition between
random matrices of order unity and the wash-out effect. As a result, the mass
spectrum for the right-handed neutrinos and the neutrino Yukawa matrix
exhibit a certain pattern, which is quite similar to that can be
understood in terms of a conventional flavor symmetry. In other words,
the flavor symmetry of the right-handed neutrino sector is {\it
  emergent} in this framework. In the case of non-thermal
leptogenesis, we find that the neutrino mass spectrum is significantly
affected in contradiction with the observations, if the inflaton mass
is smaller than the typical right-handed neutrino mass scale of
$\GEV{15}$. It suggests that the inflaton mass needs to be larger than
or comparable to $\oten{15}$\,GeV for successful non-thermal
leptogenesis.

The rest of this paper is organized as follows. In Sec.~\ref{sec:2} we
briefly review the neutrino mass anarchy hypothesis and define our notation and
framework. In Sec.~\ref{sec:3} we discuss how leptogenesis affects the
neutrino parameters. The last section is devoted for discussion and
conclusions.

\section{Neutrino mass anarchy}
\label{sec:2}
In this section we briefly review the neutrino mass anarchy and its prediction,
focusing on the see-saw mechanism.
\subsection{Preliminaries}
We consider the following see-saw Lagrangian,
\beq
{\cal L} \;\supset\; f_{ij} \, {\bar e}_{Ri} \ell_j {\bar H}+ h_{i j}\, {\bar N}_i \ell_j  H
- \frac{1}{2} M_{ij} \bar{N}_i {\bar N}_j + {\rm h.c.},
\eeq
where $i, j = \{1,2,3\}$ denote flavor indices, $\ell_i$ represents
the left-handed lepton doublets, $e_{Ri}$ are the charged lepton
singlets, $N_i$ are the right-handed neutrinos, and $H$ is the Higgs
doublet.  $f_{ij}$ and $h_{ij}$ form complex-valued $3 \times 3$
matrices of charged-lepton and neutrino Yukawa couplings, respectively, and
 $M_{ij}$ forms a complex valued $3 \times 3$ symmetric
matrix of the right-handed neutrino Majorana mass.  For later use we
also define a dimensionless matrix $X_{ij}$ as
\beq
X_{ij} \;\equiv\; \frac{M_{ij}}{M_0},
\eeq
where $M_0$ is the typical mass scale of the right-handed
neutrinos. $M_0$ can be interpreted as the B$-$L breaking scale, if the
right-handed neutrino mass arises from the vacuum expectation value
(vev) of the B$-$L Higgs boson through a renormalizable interaction
with a coupling of order unity.  We adopt $M_0 = \GEV{15}$ as a
reference value throughout this paper. Later we will briefly discuss
how our results will change for different values of $M_0$.

Quantum mechanics dictates that any states with the identical quantum
numbers should mix with each other. If there is no quantum number
which distinguishes three generations of $\ell_i$ and $N_i$, the
matrices $h_{ij}$ and $M_{ij}$ are considered to be structureless.  In
particular, they may be subject to a basis-independent random
distribution.  This is the essence of the neutrino mass anarchy.  On
the other hand, the charged lepton mass matrix (as well as that for
quarks) is probably determined by other physics such as the anthropic
considerations or conventional flavor symmetries, and so, we do not
attempt to interpret the structure of $f_{ij}$ in terms of the anarchy
here (see e.g. \cite{Donoghue:2005cf,Agashe:2008fe,Hall:2008km}).  Therefore we simply adopt a
basis where the charged-lepton Yukawa matrix is diagonalized:
\beq
{\cal L} \;\supset\; f_{\alpha} \delta_{\alpha \beta}  {\bar e}_{R\, \alpha} \ell_\beta {\bar H}+ h_{i \alpha} {\bar N}_i \ell_\alpha  H
- \frac{1}{2} M_{ij} \bar{N}_i {\bar N}_j + {\rm h.c.},
\eeq
where $\alpha$ and $\beta$ represent the lepton flavor indices,  $e,\,\mu,\,\tau$.
The anarchic nature of $h_{i\alpha}$ and $M_{ij}$ is maintained in this basis.

The neutrino Yukawa matrix $(h)_{i \alpha}$ can be diagonalized by the bi-unitary transformation,
\bea
\ell_\alpha &\rightarrow& (U_L)_{\alpha \beta}\, \ell_\beta,\\
N_i &\rightarrow & (U_R)_{ij} N_j, \\
h &\rightarrow& U_R^\dag\, h\, U_L = \left(
\bear{ccc}
h_1 & 0& 0\\
0&h_2&0\\
0&0&h_3
\eear
\right) \equiv D_h,
\label{nuY}
\eea
where $U_{L}$ and $U_R$ are unitary matrices, and we take $0 \leq h_1 \leq h_2 \leq h_3$.
Similarly, the right-handed neutrino Majorana mass matrix can be
diagonalized as
\bea
N_i &\rightarrow & (U_N)_{ij} N_j, \\
M&\rightarrow& U_N^\dag\, M\, U_N^* = \left(
\bear{ccc}
M_1 & 0& 0\\
0&M_2&0\\
0&0&M_3
\eear
\right) \equiv D_M,
\label{diagh}
\eea
where $U_N$ is a unitary matrix, and one can take $0 \leq M_1 \leq M_2
\leq M_3$ without loss of generality.

Below the scale of the right-handed neutrino, we obtain a low-energy
effective interactions containing a Majorana mass for left-handed
neutrinos:
\beq
{\cal L} \;\supset - \frac{1}{2} (m_{\nu})_{ \alpha \beta}\, \nu_\alpha \nu_\beta + {\rm h.c.},
\eeq
where
\bea
\left(m_\nu \right)_{\alpha \beta} &=&
\left( h^T X^{-1} h\right)_{\alpha \beta} \frac{v^2}{M_0}
\non\\
&=&
\left(U_L^* D_h U_R^T U_N^* D_M^{-1} U_N^\dag U_R D_h U_L^\dag \right)_{\alpha \beta}v^2,
\label{m_nu}
\eea
with $v \simeq 174$\,GeV is the vev of the Higgs field.  The light
neutrino mass can be naturally explained by the heavy right-handed
neutrino mass $M_0 \approx \GEV{15}$ close to the GUT scale in the see-saw
mechanism~\cite{seesaw}.

The neutrino mass can be diagonalized as
\beq
(m_\nu)_{\alpha \beta} \;=\; U_{MNS}^*
\left(
\bear{ccc}
m_1 & 0& 0\\
0&m_2&0\\
0&0&m_3
\eear
\right) U_{MNS}^\dag,
\eeq
where $U_{MNS}$ is a unitary matrix. There is currently no constraint
on the sign of $\Delta m^2_{31}$, but a neutrino mass spectrum with
normal hierarchy is preferentially realized in the neutrino mass anarchy,
and so, we will  assume $0 \leq m_1 \leq m_2 \leq m_3$
unless otherwise stated.

Note however that, although rare,
the inverted mass hierarchy is possible. But one should be careful when
comparing the result with the observations (\ref{mixings}) and (\ref{mass2d_obs}),
because $m_3$ is always the heaviest in our notation.
As far as the mass difference
squared is concerned, one should simply replace $m_1 \rightarrow m_3$,
$m_2 \rightarrow m_1$ and $m_3 \rightarrow m_2$ in order to compare
our results for the inverted hierarchy with the observations.
We will come back to this issue at the end of this section.

The neutrino mixing matrix $U_{MNS}$ can be parametrized as follows.
\beq
{\small
U_{MNS}\;=\; \left(
\bear{ccc}
c_{12}c_{13} & s_{12} c_{13} & s_{13} e^{-i\delta} \\
-s_{12} c_{23} -c_{12} s_{23} s_{13} e^{i\delta}&c_{12}c_{23} -s_{12}s_{23} s_{13} e^{i\delta} & s_{23} c_{13}\\
s_{12} s_{23}-c_{12}c_{23}s_{13}e^{i \delta} &-c_{12}s_{23}-s_{12}c_{23}s_{13}e^{i \delta}& c_{23}c_{13}
\eear
\right) \times {\rm diag}\left(1, \,e^{i \frac{\alpha_{21}}{2}},\,e^{i \frac{\alpha_{31}}{2}}\right),
}
\eeq
where $c_{ij} \equiv \cos \theta_{ij}$, $s_{ij} \equiv \sin
\theta_{ij}$ with $\theta_{ij} \in [0, \pi/2)$, and $\delta$,
  $\alpha_{21}$, and $\alpha_{31}$ represent the Dirac CP violation
  phase, and two Majorana CP violation phases, respectively. The CP
  phases vary from $0$ to $2 \pi$.

Lastly let us derive a relation between $U_{MNS}$ and $U_L$. We define
a unitary matrix $U_{h}$ which diagonalizes $D_h U_R^T U_N^* D_M^{-1}
U_N^\dag U_R D_h$ as
\beq
U_h^T\left(D_h U_R^T U_N^* D_M^{-1} U_N^\dag U_R D_h\right) U_h\, v^2 \;=\;
\left(
\bear{ccc}
m_1 & 0& 0\\
0&m_2&0\\
0&0&m_3
\eear
\right).
\eeq
Then $U_{MNS}$ is related to $U_L$ and $U_h$ as
\beq
U_{MNS} = U_L U_h.
\label{rel_umns}
\eeq
Note that, while $U_h$ depends on the mass and mixing of the
right-handed neutrinos, it is independent of the mixing of the lepton
doublets, $U_L$.  This relation is important when we consider
leptogenesis.

\subsection{Random matrix and measure}
The neutrino mass anarchy assumes the basis-independent random
distribution of the matrices $h$ and $M$~\cite{Haba:2000be}.  Here we
quote some of the results in Ref.~\cite{Haba:2000be} without
derivation.

Let us start with how to obtain a basis-independent random matrix,
$(h)_{ij}$, with each element of order unity. We may
generate a random number $z$ for each element,  uniformly distributed in a region of $-1
\leq {\rm Re}[z] \leq 1$ and $-1 \leq {\rm Im}[z] \leq 1$. However,
thus generated matrix is not basis-independent, as it changes its form
under the $U(3)$ rotation of the generations.  In order to obtain a
basis-independent random matrix, we need to impose ${\rm Tr}[h h^\dag]
\leq 1$, which makes the distribution invariant under $U(3)$. Similarly we can
generate a random symmetric matrix $X$.

Next question is the distribution of the eigenvalues $h_1, h_2$, and
$h_3$, and the mixing matrices $U_L$ and $U_R$, which should be
invariant under U(3).  The invariant measure of $h$ is given by
\beq
d h \;=\; F_D(h_1,h_2,h_3)\, \prod_{i=1}^{3} d h_i\, \frac{dU_L dU_R}{d \varphi}
\eeq
with
\beq
F_D(h_1,h_2,h_3) \;\equiv\;
(h_1^2-h_2^2)^2 (h_2^2-h_3^2)^2 (h_3^2-h_1^2)^2 h_1 h_2 h_3,
\eeq
where $dU_L$ and $dU_R$ represent the Haar measure of $U_L$ and $U_R$,
respectively. The $d \varphi$ in the denominator mods out the three
redundant phases; we can see this by noting that the decomposition
(\ref{diagh}) is not unique, and it is invariant under multiplication
of $U_L$ and $U_R$ by a diagonal unitary matrix.  Similarly, the
measure of $M$ is given by
\beq
d M \;=\; F_M(M_1,M_2,M_3) \, \prod_{i=1}^{3} d M_i\, dU_N
\label{FM}
\eeq
with
\beq
F_M(M_1,M_2,M_3) \;\equiv\; (M_1^2-M_2^2) (M_2^2-M_3^2) (M_3^2-M_1^2) M_1 M_2 M_3,
\eeq
where $dU_N$ represents the Haar measure of $U_N$.

Note that, while the distributions of the mixing angles are determined
uniquely by the U(3) invariance, the measure of the eigenvalues can,
in general, depend on an additional factor that is invariant under
$U(3)$, such as ${\rm Tr}[h h^\dag]$ or ${\rm det}[h]$.  Throughout
this paper we assume that there is no such additional factor in the
measure; this is called ``linear measure" in the literature. Later we will briefly
comment on how our results may change if other measure is adopted.

Lastly let us give the measure of the neutrino mass matrix in the see-saw mechanism.
The linear measure is given by
\bea
dh \, dM & \propto & F_D(h_1,h_2,h_3) F_M(M_1,M_2,M_3) \, \prod_{i=1}^{3} d h_i \prod_{j=1}^{3} dM_j\,
\frac{dU_L dU_{NR}}{d \varphi},
\eea
where $dU_{NR}$ is the Haar measure of $U_{NR} \equiv U_N^\dag U_R$.
The neutrino mass eigenvalues can be obtained by diagonalizing
$m_\nu =U_L^* D_h U_{NR}^T  D_M^{-1} U_{NR} D_h U_L^\dag\,v^2$.
We may replace $dU_L$ with $dU_{MNS}$ by using \REF{rel_umns}.
Thus,  we can see that the mixing angles obey the invariant
Haar measure of $U_{MNS}$, since there is no way to distinguish three generations. It is given by
\beq
dU_{MNS} \;\propto\; d s_{12}^2 dc_{13}^4 ds_{23}^2\, d \delta\, d \alpha_{21} d \alpha_{31}.
\label{dUMNS}
\eeq
From \REF{rel_umns} we can see that the Haar distribution of $U_{MNS}$
arises from the U(3)-invariance of $U_L$. Thus, even if the
distribution of $U_h$ is significantly distorted by e.g. leptogenesis,
the Haar distribution of the $U_{MNS}$ matrix remains intact.

 For practical purposes, it is useful to
parametrize the light neutrino mass matrix as
\beq
m_\nu \;=\; G^T D_M^{-1} G\, v^2,
\label{combined}
\eeq
where $M_i$ obeys the linear measure distribution $F_M(M_1,M_2,M_3)$
and $G$ is a complex valued $3 \times 3$ random matrix generated as
explained above.  In this method, the neutrino Yukawa couplings of the
right-handed neutrino mass eigenstates are directly obtained as the matrix $G$,
and so, it is convenient when we discuss leptogenesis.

\begin{figure}[t!]
\begin{center}
\includegraphics[width=6cm]{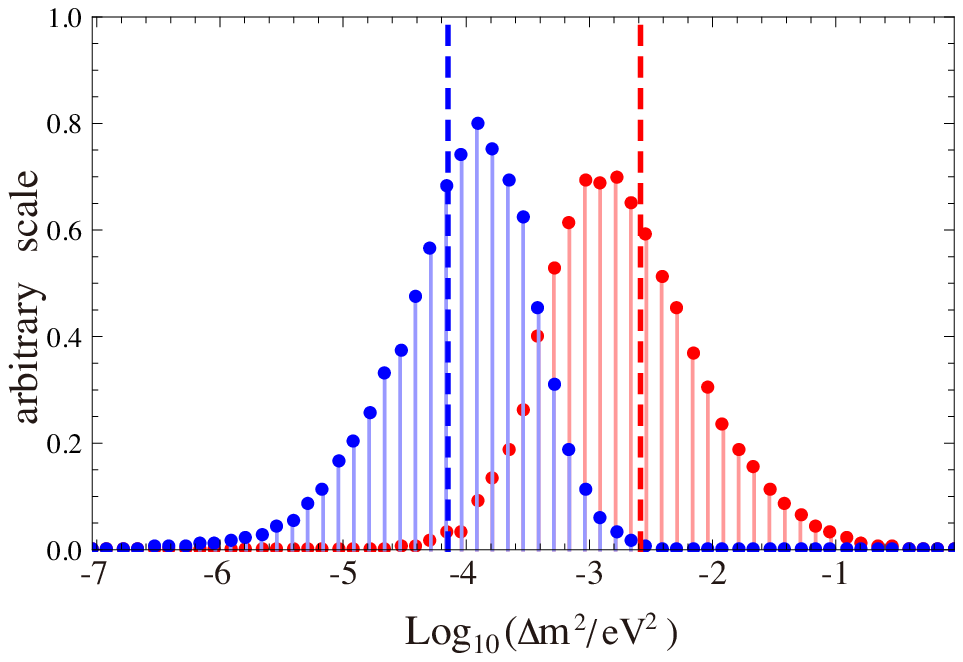}\qquad
\includegraphics[width=6cm]{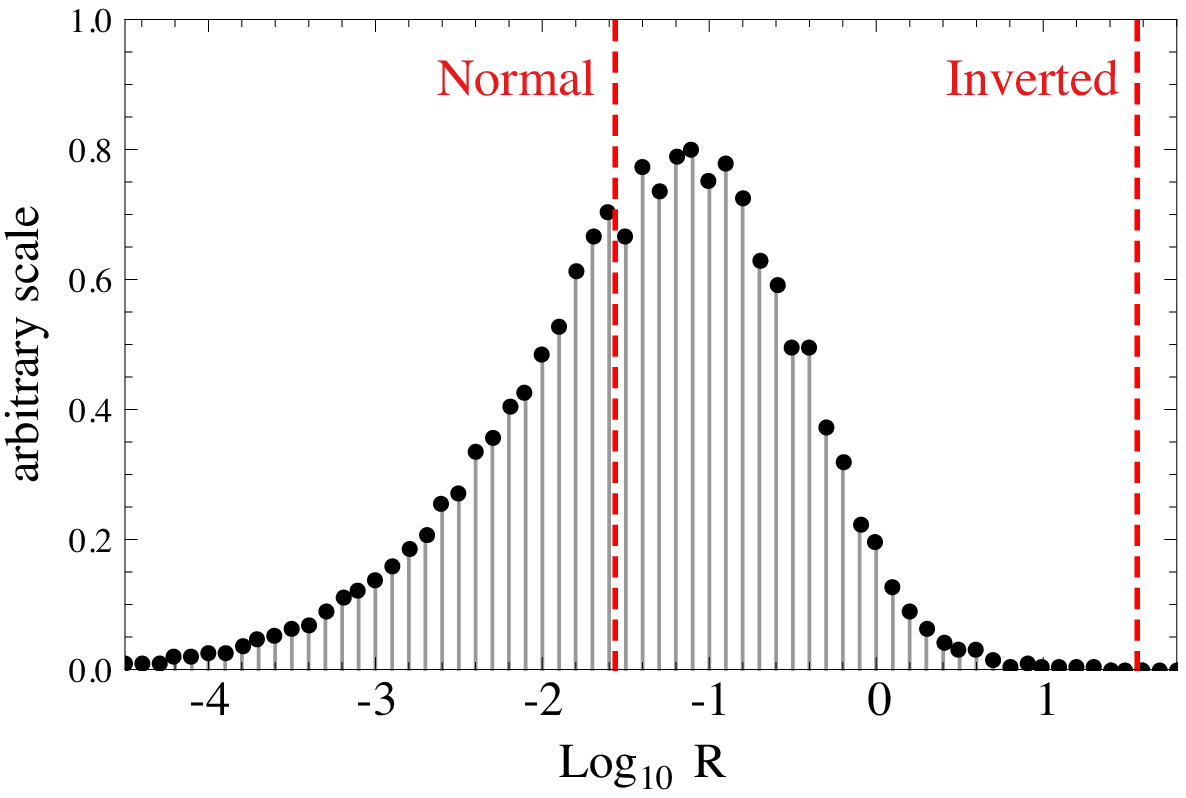}
\vskip 0.5cm
\includegraphics[width=6cm]{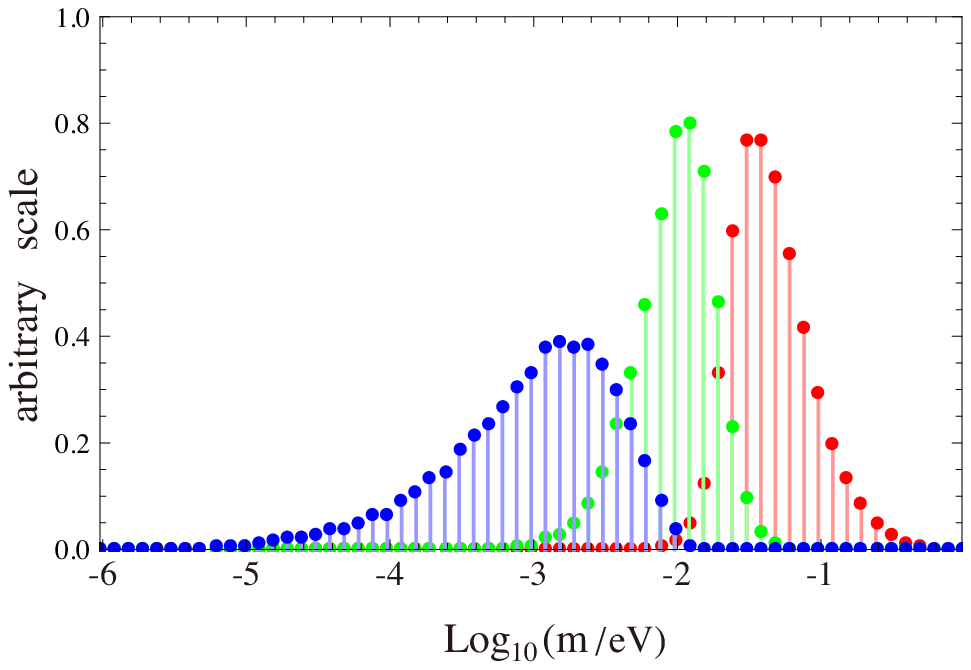}\qquad
\includegraphics[width=6cm]{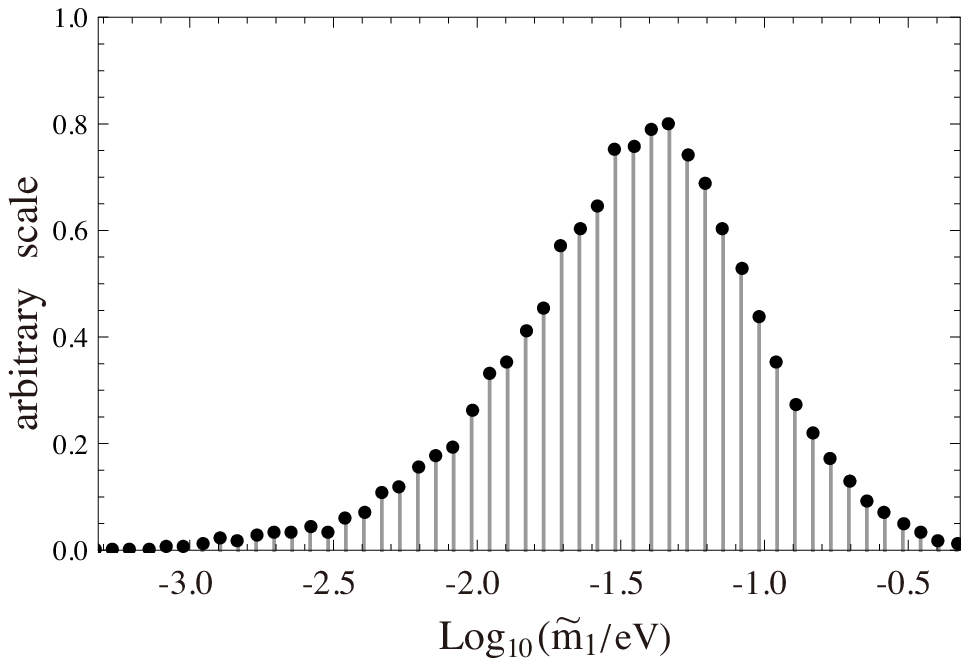}
\vskip 0.5cm
\includegraphics[width=6cm]{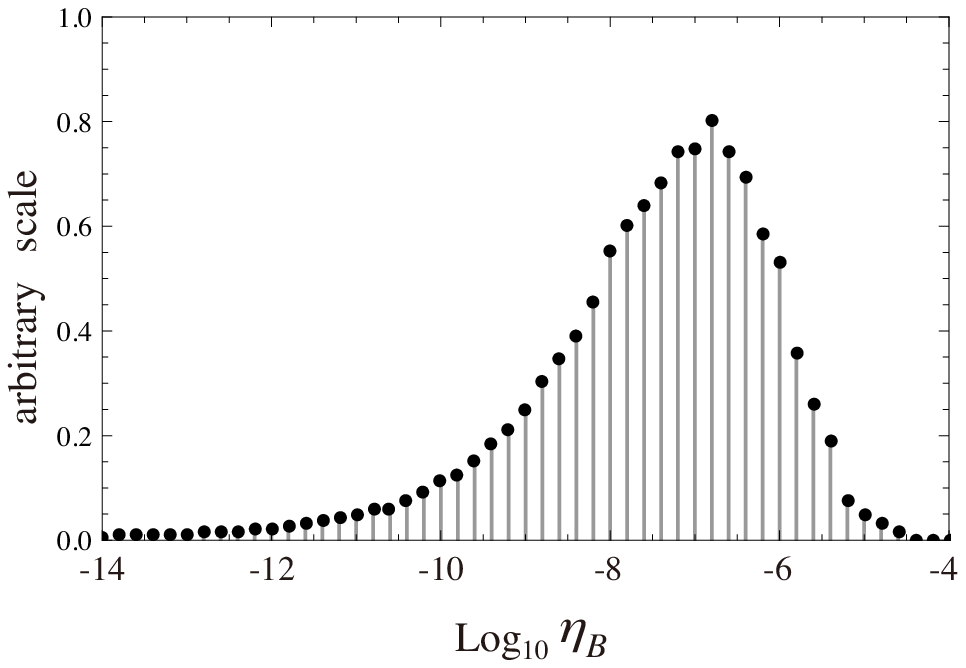}\qquad
\includegraphics[width=6cm]{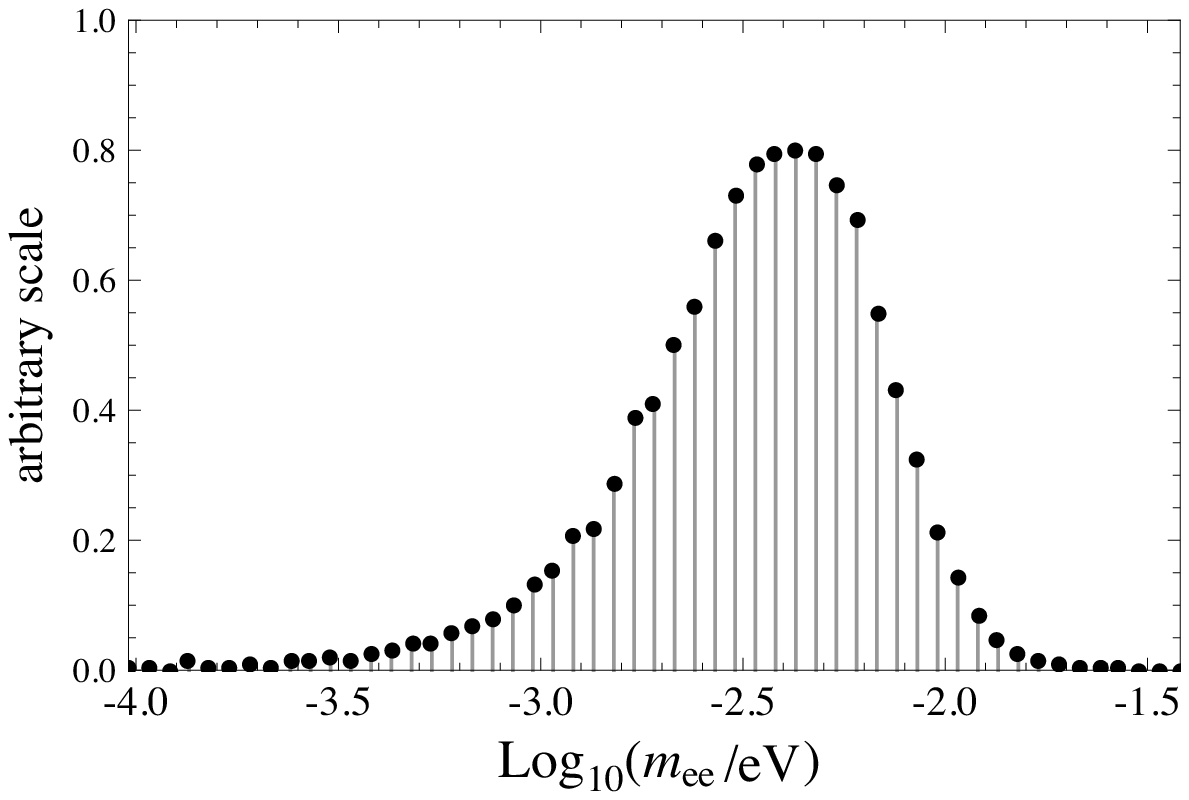}
\caption{ The distributions of the mass squared differences $(\Delta
  m^2_{32} \,({\rm red}), \Delta m^2_{21}\,({\rm blue}))$, their ratio
  $R$, the neutrino masses $(m_1\,({\rm blue}), m_2 \,({\rm green}),
  m_3\,({\rm red}))$, ${\widetilde m}_1$ (to be defined in Eq.~(\ref{m1tilde})),
   the baryon-to-photon ratio $\eta_B$, and $m_{ee}$ (to be defined in Eq.~(\ref{mee})).  The
  vertical lines represent the observed mass squared differences
  (\ref{mass2d_obs}), $R \approx 0.03$ (normal hierarchy) and $R \approx 30$ (inverted hierarchy).  }
\label{fig:dist}
\end{center}
\end{figure}

Since the light neutrino mass matrix is obtained by the product of
several random matrices, its mass eigenvalues exhibit a mild
hierarchy.  In particular, the ratio of mass squared difference, $R
\equiv (m_2^2-m_1^2)/(m_3^2-m_2^2)$, can be naturally as small as the
observed value $\sim 1/30$ for the normal hierarchy. See Fig.~\ref{fig:dist} for the
distributions of the mass squared differences $(\Delta m^2_{32},
\Delta m^2_{21})$, $R$, $(m_1, m_2, m_3)$, ${\widetilde m}_1$ (to be defined in Eq.~(\ref{m1tilde})),
  the baryon-to-photon ratio $\eta_B$, and $m_{ee}$ (to be defined in Eq.~(\ref{mee})).  Here we have assumed
that the typical values of $h_{i \alpha}$ and $X_{ij}$ are of order
unity, and imposed the constraints ${\rm Tr}[h h^\dag] \leq 1$ and
${\rm Tr}[X X^\dag] \leq 1$, and fixed $M_0$ to be $\GEV{15}$.  One
can see from the figure that the distributions are consistent with the
observation.  In particular, the mild hierarchy $R \sim 1/30$ is
nicely explained. Note that the distribution of $R$ remains intact even
if we change the typical values of the neutrino Yukawa matrix and $M_0$, while
the neutrino mass distribution is affected. This is no longer the case
when leptogenesis is taken into account, and both $R$ and $m_i$ sensitively
depend on the reheating temperature.

We can see from the figure that
the normal hierarchy is preferred since $\Delta  m^2_{32}$ tends to be
larger than $\Delta  m^2_{21}$. In our notation, the inverted hierarchy
is realized if $\Delta  m^2_{32} \ll \Delta  m^2_{21}$, and in order to be
consistent with the observations,  the distribution of $\Delta m_{32}^2$ and $\Delta m_{21}^2$
should overlap with the left (blue) and right (red) dashed vertical lines, respectively.
We can see that, although rare, the inverted hierarchy is indeed possible in the
neutrino mass anarchy hypothesis.

We will study how the distributions are affected if we impose the
leptogenesis in the next section.

\section{Leptogenesis and anarchy}
\label{sec:3}
We are interested in the conditional distribution of the neutrino
parameters where leptogenesis works successfully.  In the leptogenesis
scenario, the lepton asymmetry is generated by the out-of-equilibrium
decay of right-handed neutrinos.  In the thermal leptogenesis scenario
with zero initial abundance, the right-handed neutrino is generated
thermally by inverse decay and scattering processes in thermal
plasma. On the other hand, the right-handed neutrino is generated
non-thermally by the inflaton decay in the non-thermal leptogenesis
scenario~\cite{Asaka:1999jb,Hamaguchi:2001gw}.  We will see that this
distinction is crucial in the neutrino mass anarchy. To simplify our
analysis, we focus on a case in which the final lepton asymmetry is
predominantly generated by the decay of the lightest right-handed
neutrino, $N_1$, and we do not consider effects of the flavored
leptogenesis~\cite{Abada:2006fw,Nardi:2006fx}.\footnote{We adopt a basis in which the right-handed neutrino
mass matrix is diagonalized in this section. See discussion below \REF{combined} for how we generate
random matrices.} The resonant leptogenesis~\cite{Pilaftsis:1997jf} is disfavored in the neutrino mass anarchy, because the
  measure (\ref{FM}) forces the right-handed neutrino masses to be
  apart from each other.

As is well known, the reheating temperature after inflation is bounded
below, $T_R \gtrsim 10^9 (10^6)$\,GeV, in order for (non-)thermal
leptogenesis to account for the observed baron asymmetry~Ref.~\cite{Buchmuller:2005eh}.  Thus the
reheating temperature $T_R$ is an important input parameter for
leptogenesis, but its precise value is poorly known, and so, we treat
$T_R$ as a free parameter and see how the distribution of the neutrino
parameters changes as we vary $T_R$. We do not take account of the prior
distributions of $T_R$ and the resultant baryon asymmetry, because
they are likely distorted by the anthropic conditions if leptogenesis
is responsible for the origin of matter.  On the other hand, if both
the neutrino mass anarchy and leptogenesis are realized in nature, the
observed neutrino mass squared differences and the mixing angles
should be {\it typical} in the conditional distribution (as long as
the light neutrino masses are irrelevant to the existence of life).

\subsection{Preliminaries}
\label{sec:3-0}
 The decay rate of $N_1$ at tree
level is given by
\beq
\Gamma_1(N_1 \rightarrow H+\ell_\alpha) = {\bar \Gamma}_1(N_1 \rightarrow H^\dag + \ell_\alpha^\dag)
= \frac{1}{16 \pi}\left(h h^\dag\right)_{11} M_1,
\eeq
ignoring the masses of the final states.
The CP asymmetry $\varepsilon_1$ of the decay of $N_1$ reads~\cite{Flanz:1994yx,Covi:1996wh,Buchmuller:2003gz}
\beq
\varepsilon_1\; = \; \frac{1}{8 \pi} \frac{1}{\left(h h^\dag\right)_{11}} \sum_{j=2,3} {\rm Im}\left[\left(h h^\dag\right)^2_{1j}\right]
f\left(\frac{M_j}{M_1}\right)
\eeq
with
\beq
f(x)\;\equiv\; x(1 + x^2) \log{(1 + x^{-2})} - x + \frac{x}{x^2-1}.
\eeq
Then, the final baryon-to-photon ratio $\eta_B$ is given by
\beq
\eta_B\;\approx\;3 \frac{a_{sph}}{g_*} \varepsilon_1 \kappa,
\eeq
where $a_{sph}$ represents the sphaleron conversion factor, and it is
equal to $29/78$ in the SM, $g_* \approx 100$ counts the relativistic
degrees of freedom at the $N_1$ decay, and $\kappa$ denotes the
efficiency factor~\cite{Buchmuller:2005eh}. We take account of both
$\Delta L =1$ and $\Delta L =2$ wash-out processes in our analysis.
For later use, let us also define the following parameter:
\beq
{\widetilde m}_1 \;\equiv\; \frac{\left(h h^\dag\right)_{11} v^2}{M_1},
\label{m1tilde}
\eeq
which is proportional to the ratio of the decay rate of $N_1$ to the Hubble
parameter at the temperature $T = M_1$.

\subsection{Invariant Haar measure of $U_{MNS}$}
\label{sec:3-1}
\begin{figure}[t!]
\begin{center}
\includegraphics[width=15cm]{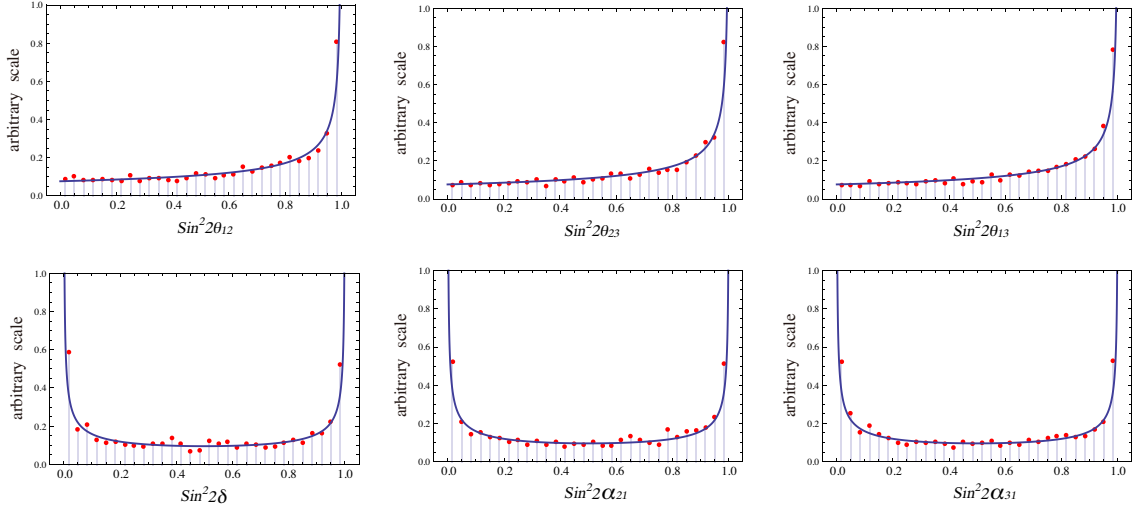}
\caption{The solid lines represent the Haar measure distribution of the mixing angles and CP phases given by \REF{dUMNS}. The red dots represent
the distributions when we impose a successful leptogenesis constraint: $M_{1, max} = \GEV{13}$ and $10^{-10} \leq \eta_B \leq 10^{-9}$.
}
\label{fig:haar}
\end{center}
\end{figure}

Here we show that the distribution of $U_{\rm MNS}$ is orthogonal to
the parameters relevant for leptogenesis. Since the lepton asymmetry
is generated by the decay of $N_1$, it is $M_1$ and $h_{1\alpha}$ that
are especially relevant for leptogenesis.  
As we have seen, the neutrino Yukawa matrix $(h)_{i \alpha}$ always appears in a form of
$(h h^\dag)$ in the decay rate, the CP asymmetry, and the efficiency
parameter.  Since $(h h^\dag)$ is invariant under the U(3) rotation of
$\ell_\alpha$ (see \REF{nuY}), $U_L$ is not constrained by the
leptogenesis.  Whatever constraints on $M_1$ and/or $(h h^\dag)$ are
imposed, it does not affect the distribution of $U_{L}$. (The
distribution of $U_h$ is affected.)  Considering the relation between
$U_{MNS}$ and $U_L$ given by \REF{rel_umns} and the translational
invariance of the Haar distribution, we conclude that the distributions
of the neutrino mixing angles as well as Dirac and Majorana CP
violation phases are independent of leptogenesis.  Thus, the measure of
$U_{MNS}$ is still given by \REF{dUMNS} even if we require successful
leptogenesis. We have confirmed this property numerically as well.
In Fig.~\ref{fig:haar}, one can see that the mixing angles and the CP violation
phases are determined by the Haar measure of U(3).
 This is a good news: the success of the neutrino mass
anarchy about the mixing angles is maintained, even if we require the
successful leptogenesis.

\subsection{Thermal leptogenesis}
\label{sec:3-2}
In thermal leptogenesis with zero initial abundance, $N_1$ is
thermally produced by the inverse decay and scattering processes, and
its out-of-equilibrium decay generates lepton asymmetry, which is
converted to the baryon asymmetry.
Requiring successful thermal leptogenesis affects the original
distribution of the neutrino parameters in two ways. First we consider
the effect of the reheating temperature $T_R$.  To simplify our
argument, we assume that the radiation dominated Universe started with
temperature $T_R$, and there was no thermal plasma before the
reheating. In the case of the usual exponential decay, there is
thermal plasma even before the reheating. However, even if the right-handed neutrinos
are produced before reheating, they will be diluted
by the subsequent entropy production.  Such a crude approximation is
sufficient for our purpose.

If $T_R$ is close to $M_0(\equiv 10^{15}\,{\rm GeV})$, all the three right-handed neutrinos
are thermalized, and the leptogenesis is possible. The distribution of the baryon asymmetry
is shown in Fig.~\ref{fig:dist}. The conditional distribution is obtained if we limit ourselves to the neutrino
parameters leading to the correct amount of the baryon asymmetry.

If $T_R \ll M_0 $, on the other hand,  the three right-handed
neutrinos tend to be too heavy to be produced.  So, for most of the
neutrino parameters, thermal leptogenesis does not work. However,
although rare, the lightest right-handed neutrino can be light enough,
by chance, to be thermally produced. So, successful leptogenesis is
possible only in such subset ${\cal S}_1$ satisfying $M_1 \lesssim z\,
T_R$:
\beq
{\cal S}_1:\,\,M_1 \lesssim z \,T_R \ll M_{2} \leq M_3,
\label{s1}
\eeq
where $M_2$ and $M_3$ are comparable to $M_0$.  In the weak washout
regime, $z \approx 1$, while $z$ is about $4 - 6$ in the strong
washout regime~\cite{Buchmuller:2005eh}.  In ${\cal S}_1$, the
distribution of $M_1$ is peaked at $M_1 \sim z T_R$.  It is clear that
the neutrino mass distribution in ${\cal S}_1$ is far from the
observed one.  That is to say, one of the light neutrino will be much
heavier than the sub-eV scale. This can be understood by noting that
the right-handed neutrino mass appears in the denominator in the
see-saw formula for the light neutrino mass \REF{m_nu}.  Therefore, as
long as the neutrino Yukawa couplings are of order unity, one of the
light neutrino masses will become much heavier than the other two.

In order to avoid this problem, the neutrino Yukawa couplings, $h_{1
  \alpha}$, must be suppressed.  Such suppression of the $M_1$ and
$h_{1 \alpha}$ can be easily realized by a simple U(1) flavor symmetry
under which $N_1$ is charged. However, as we are considering the
neutrino mass anarchy without {\it any} flavor symmetries in the
neutrino sector, we need some other explanation. Intriguingly, such
suppression is actually {\it required} in thermal leptogenesis.  This
is because of the wash-out effect: the neutrino Yukawa coupling $h_{1
  \alpha}$ must be suppressed because otherwise the resultant lepton
asymmetry would be erased efficiently.  Thus, the successful
leptogenesis selects the following subset:
\beq
{\cal S}_2:\,\, \sum_\alpha |h_{1 \alpha}|^2 \ll 1.
\label{s2}
\eeq
This is the second effect of thermal leptogenesis.  The analytic and
numerical estimate of the upper bound can be found in
e.g. Ref.~\cite{Buchmuller:2005eh}.  We will see that the upper bound
can be expressed in terms of ${\widetilde m}_1$ as ${\widetilde m}_1
\lesssim {\cal O}(0.1)$\,eV. (Note that we have not imposed the
observed mass squared difference.)

To summarize, successful thermal leptogenesis selects the subset
${\cal S}_1 \bigcap {\cal S}_2$, where the hierarchical mass spectrum
for the right-handed neutrinos, $M_1 \ll M_{2}, M_3$, as well as the
suppressed neutrino Yukawa couplings, $|h_{1 \alpha}| \ll 1$, are
realized.  Such feature is genuinely emergent from the anarchy and
thermal leptogenesis: no flavor symmetry is required.

\vspace{5mm} Before proceeding, let us describe our strategy. If the
neutrino mass anarchy and the leptogenesis are indeed realized in
nature, the observed neutrino parameters \REF{mixings} and
\REF{mass2d_obs} should be typical ones in the conditional
distribution.  Since the mixing angles obey the Haar distribution,
we will focus on the distribution of the neutrino masses.  To this end,
we will study the conditional distribution of the neutrino masses for which thermal
leptogenesis works successfully with given $T_R$.  Note that we do not
impose the observed values \REF{mass2d_obs}.  If the observed mass
squared differences are not typical in the obtained distribution, we
conclude that such a framework is disfavored; either the true reheating
temperature should be different, or the assumptions of the neutrino
mass anarchy and/or thermal leptogenesis are wrong. In this paper, we
do not estimate the goodness of fit, since the qualitative
understanding is fully adequate for our purposes. We leave a detailed
statistical test for future work.

\begin{figure}[t!]
\begin{center}
\includegraphics[width=6cm]{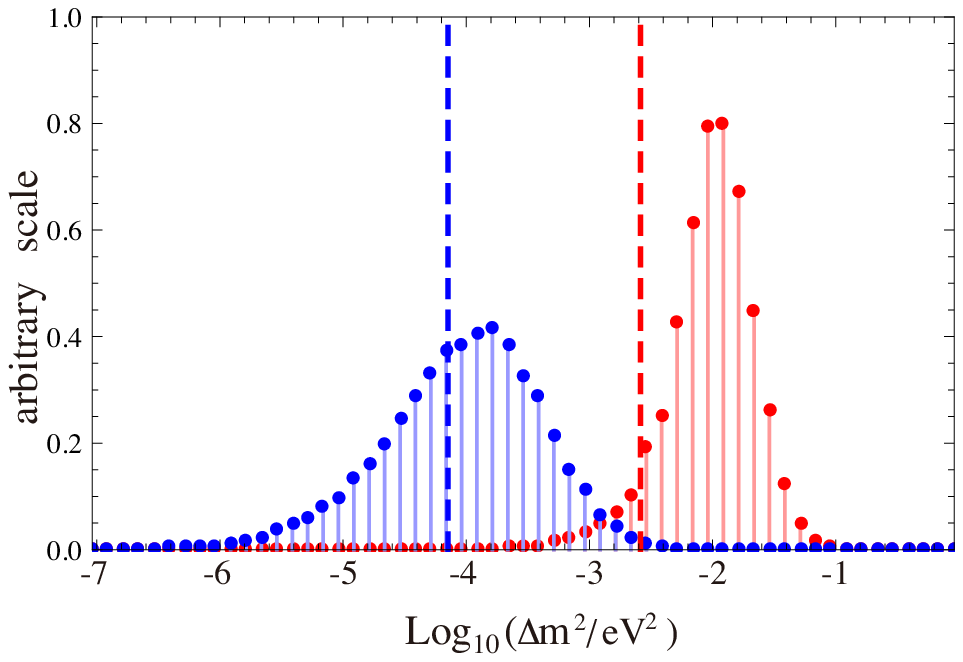}\qquad
\includegraphics[width=6cm]{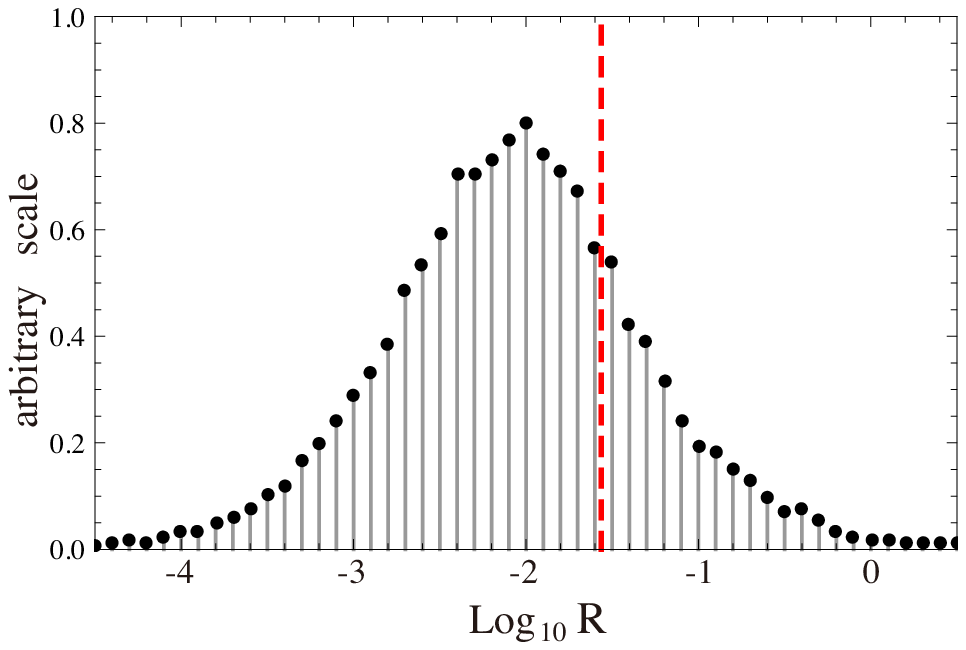}
\vskip 0.5cm
\includegraphics[width=6cm]{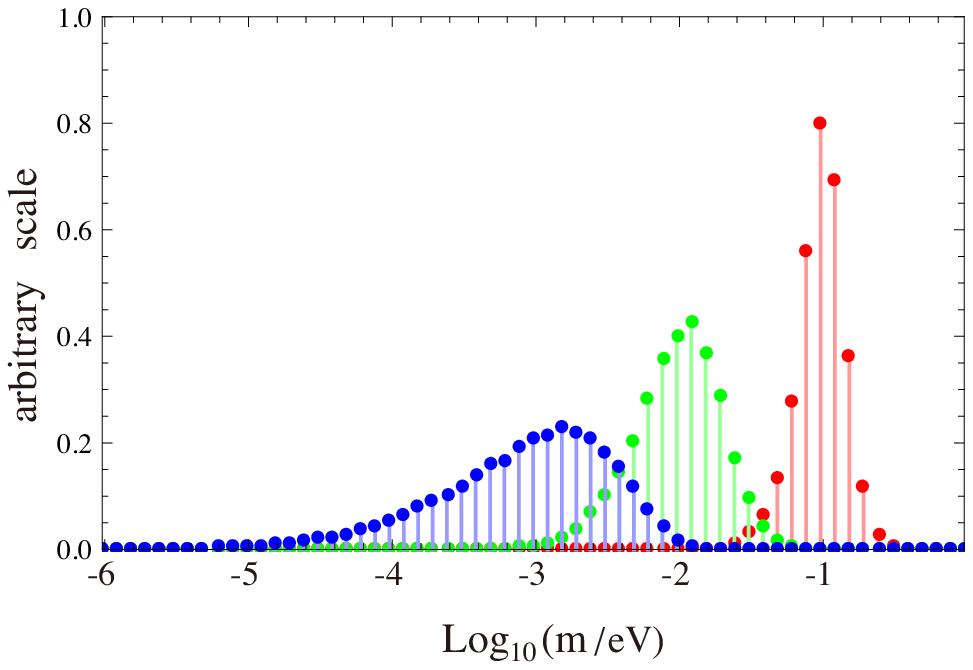}\qquad
\includegraphics[width=6cm]{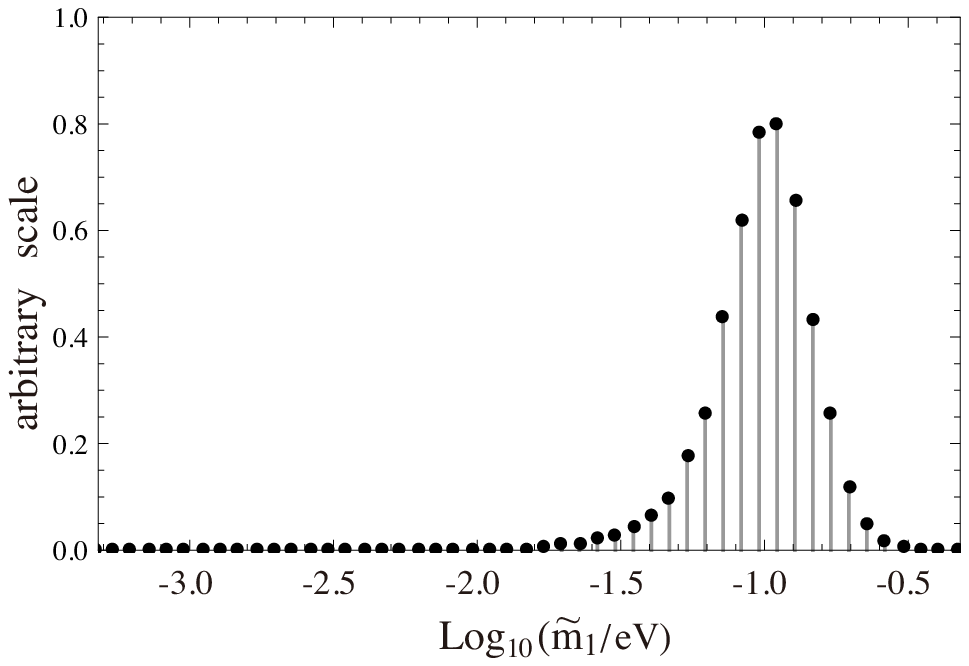}
\caption{
Same as Fig.~\ref{fig:dist}, except that we have required successful leptogenesis \REF{suclep}.
No constraint on $M_1$ is imposed.
 }
\label{fig:dist_1e15}
\end{center}
\end{figure}

Now let us go into details and see how the distributions change as we
vary $T_R$.  In Fig.~\ref{fig:dist} we have shown the original
distribution. Now we impose the successful leptogenesis; namely, we
require the baryon asymmetry to be in the following range:\footnote{
  The reason why we do not impose the observed value, $\eta_B = (6.19
  \pm 0.15) \times 10^{-10}$~\cite{Komatsu:2010fb}, is to increase the
  number of random matrices satisfying the above criterion. Our main
  purpose here is to obtain the qualitative understanding of how the
  distributions of the neutrino parameters are modified by imposing
  successful leptogenesis.   }
\beq
5\times 10^{-10} \leq \eta_B \leq 7\times 10^{-10}.
\label{suclep}
\eeq
In Fig.~\ref{fig:dist_1e15} we show the distribution of the mass
squared differences $(\Delta m^2_{32}, \Delta m^2_{21})$, $R$,
$(m_1, m_2, m_3)$, and $\tilde m_1$. We have generated one million
sets of random matrices satisfying
\REF{suclep}.  Note that here we do not impose any constraint on
$M_1$. This corresponds to the case of a high reheating temperature,
$T_R \sim M_0$. We can see that the typical value of $R$ decreases by
one order of magnitude, compared to the original distribution in
Fig.~\ref{fig:dist}.  This can be understood as follows. Since the
original distribution of the baryon asymmetry is peaked around
$\eta_B \sim 10^{-7}$ (see Fig.~\ref{fig:dist}), we need to suppress the baryon asymmetry to
satisfy \EQ{suclep}. This can be achieved by decreasing the value of
$M_1$, which leads to the increase of ${\widetilde m}_1$. (Note that
the value of ${\widetilde m}_1$ is bounded above, ${\widetilde m}_1
\lesssim {\cal O}(0.1){\rm\,eV}$, for the successful leptogenesis.)
As a result, the mass of the heaviest neutrino, $m_3$, tends to be
heavier, suppressing $R$.  We can see that the distribution of $\Delta
m_{32}^2$ is in tension with the observations.

\begin{figure}[t!]
\begin{center}
\includegraphics[width=6cm]{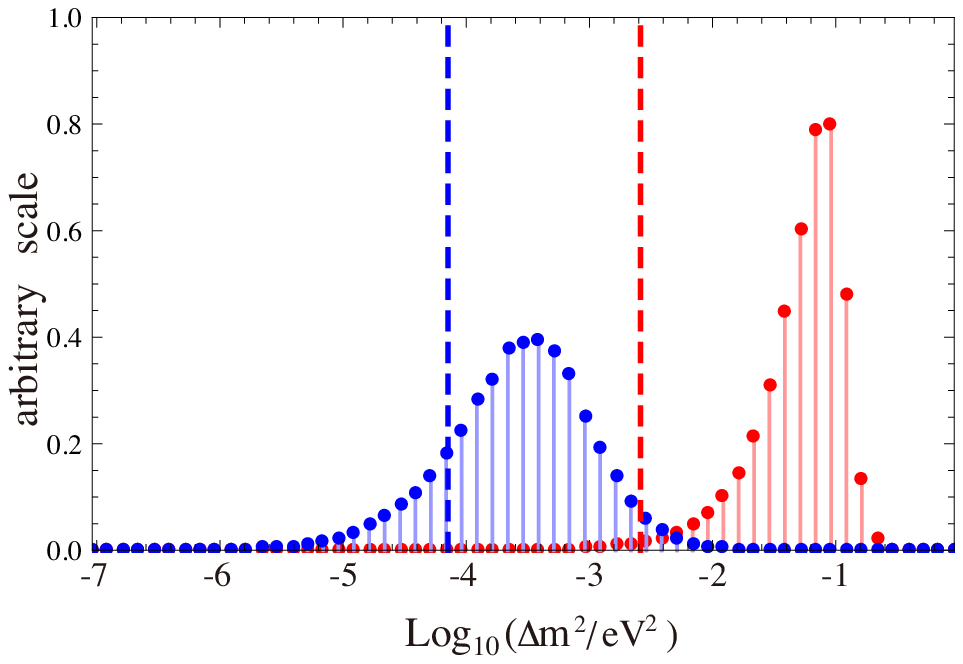}\qquad
\includegraphics[width=6cm]{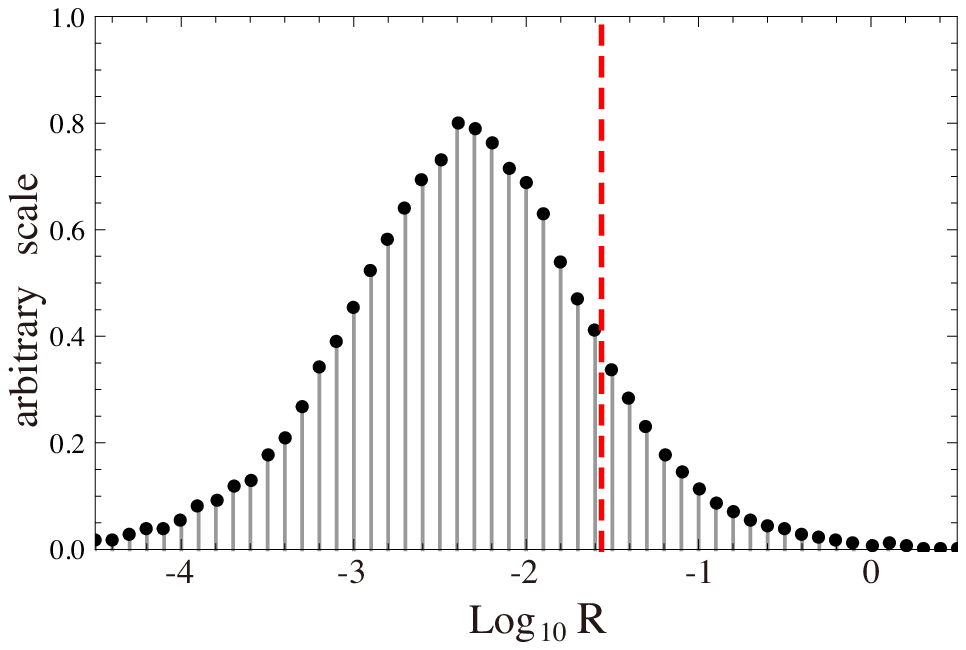}
\vskip 0.5cm
\includegraphics[width=6cm]{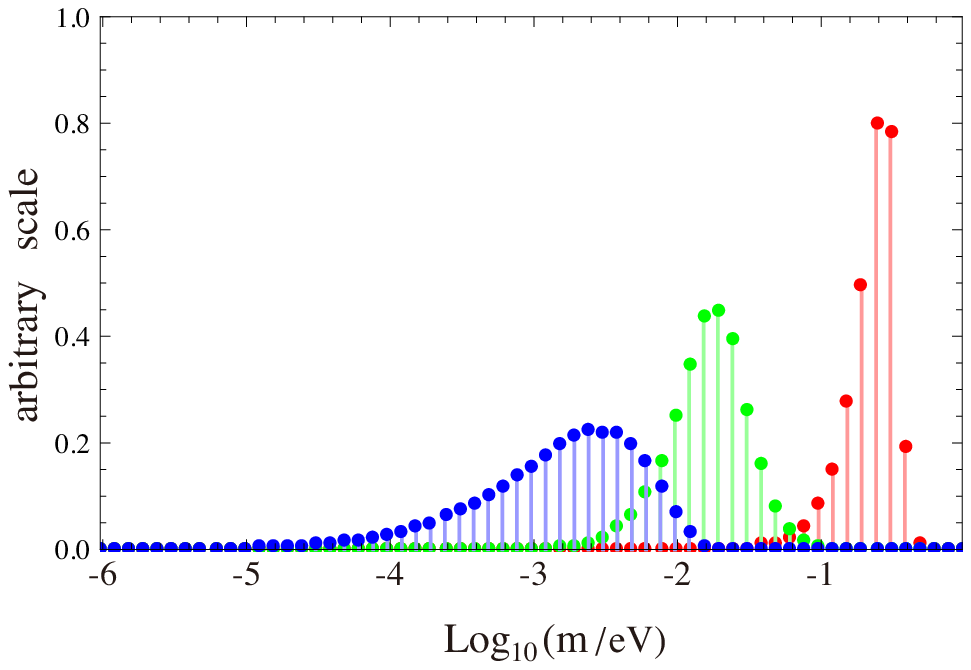}\qquad
\includegraphics[width=6cm]{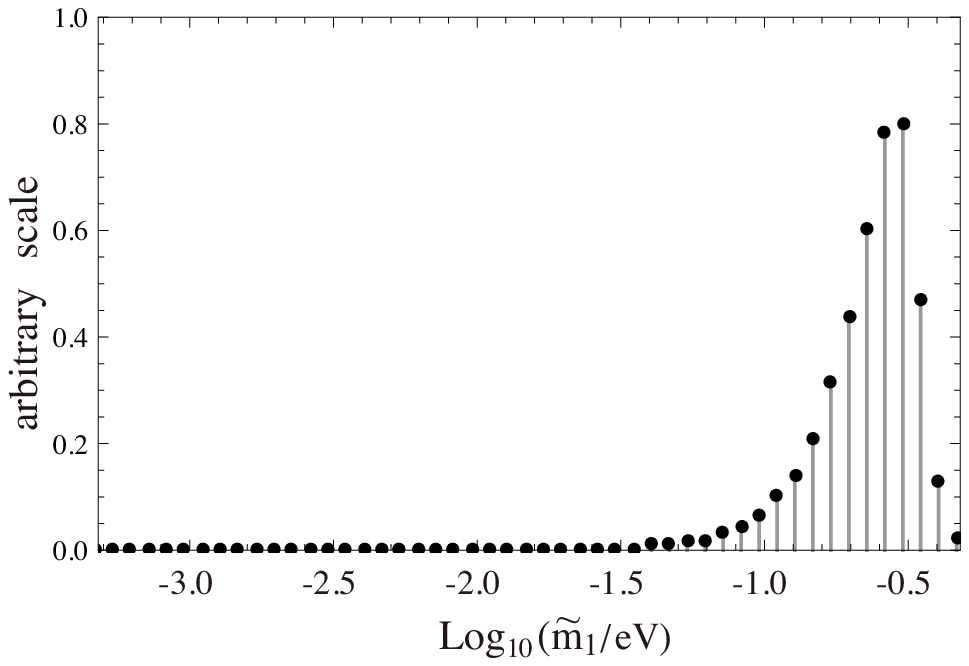}
\caption{
Same as Fig.~\ref{fig:dist_1e15}, except that we have imposed a constraint $M_1 \leq \GEV{13}$.
 }
\label{fig:dist_1e13}
\end{center}
\end{figure}

\begin{figure}[t!]
\begin{center}
\includegraphics[width=6cm]{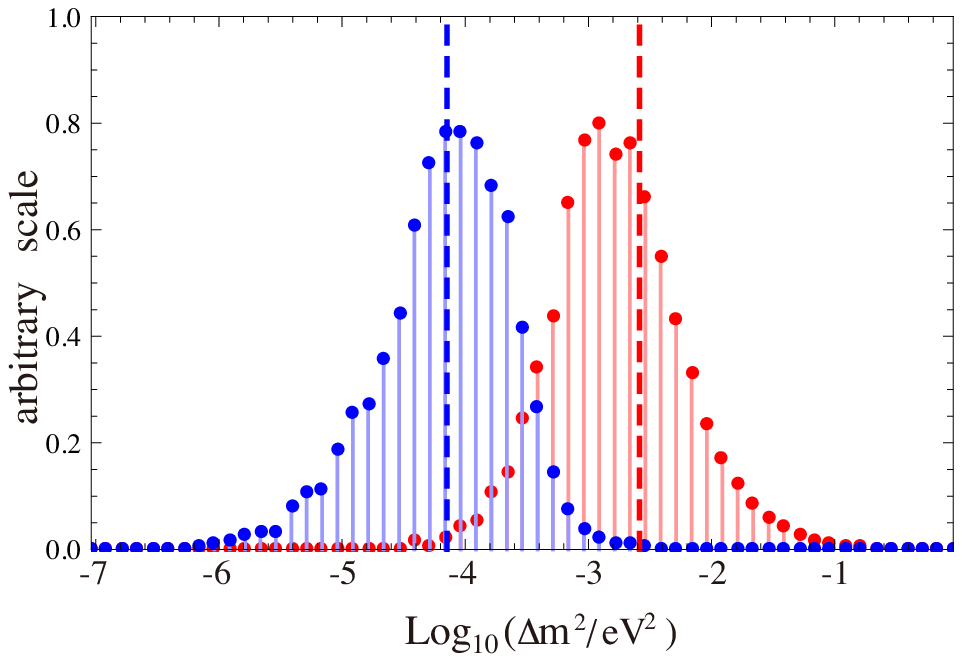}\qquad
\includegraphics[width=6cm]{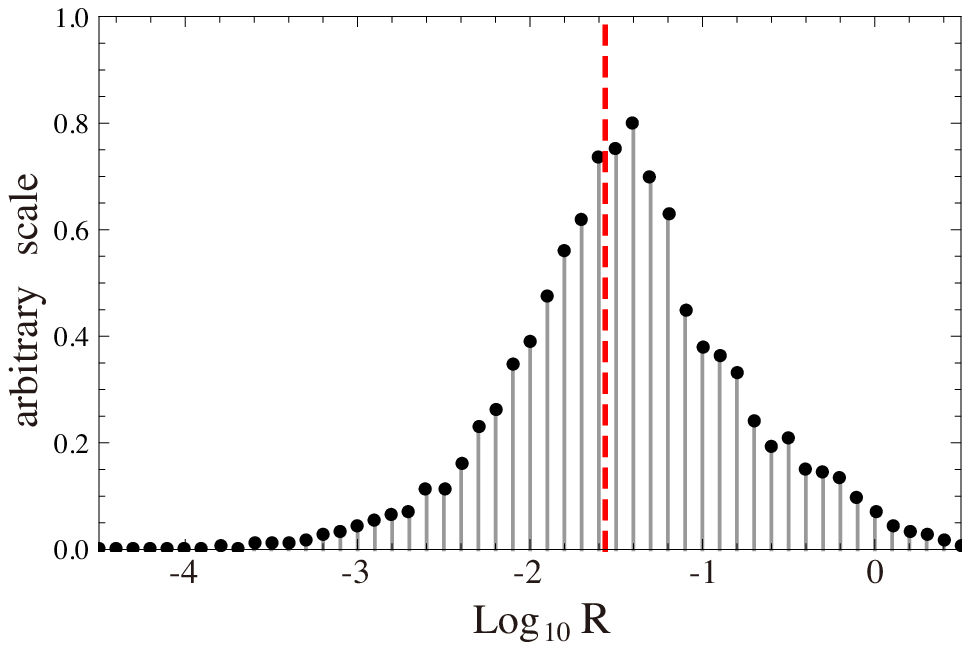}
\vskip 0.5cm
\includegraphics[width=6cm]{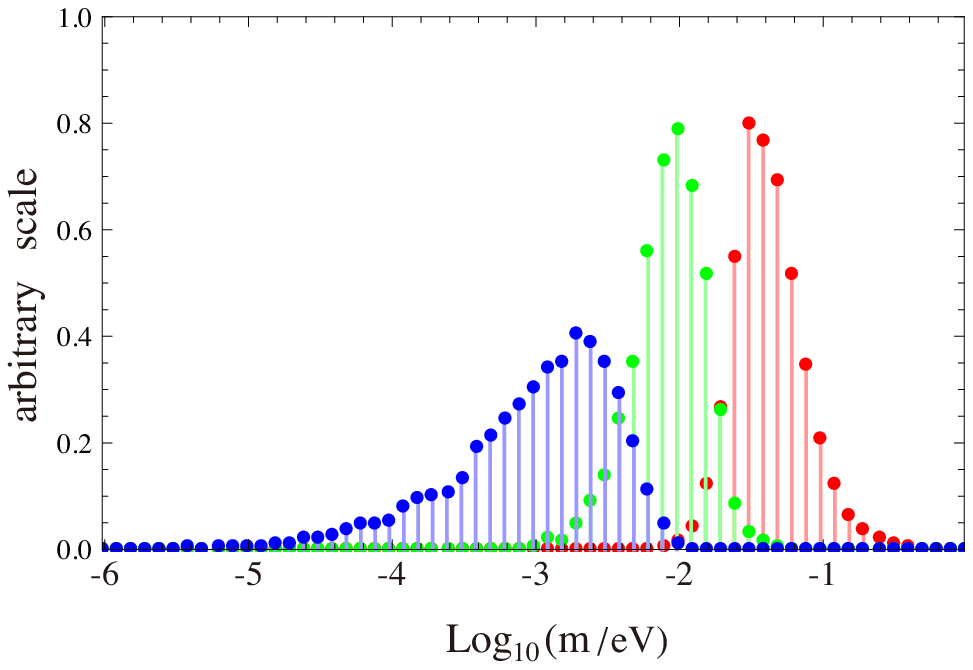}\qquad
\includegraphics[width=6cm]{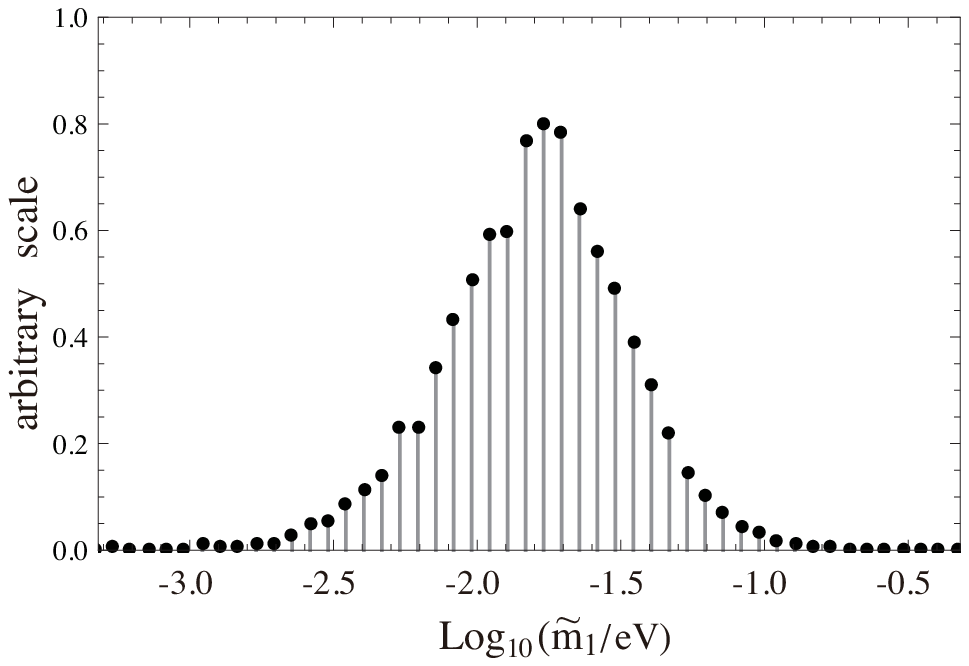}
\includegraphics[width=6cm]{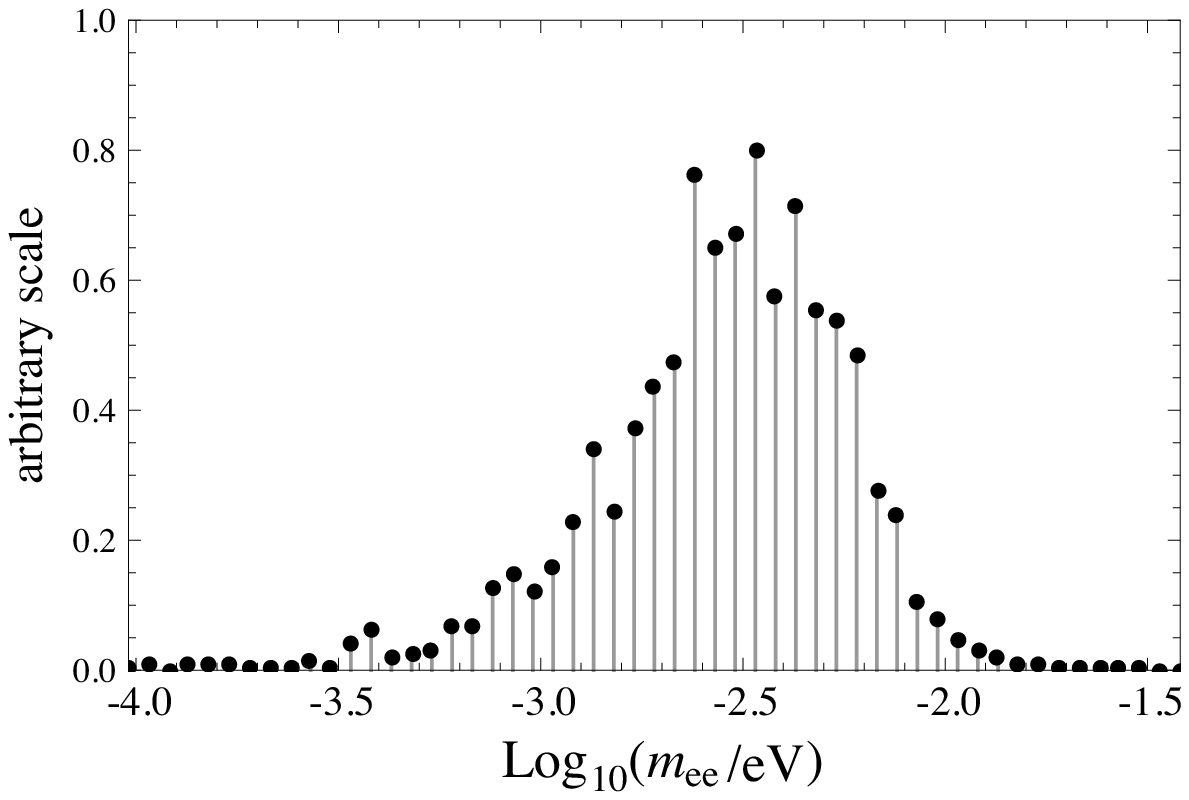}
\caption{
Same as Fig.~\ref{fig:dist_1e15}, except that we have imposed a constraint
$M_1 \leq 5\times 10^{10}$\,GeV. The distribution of $m_{ee}$ is also shown.
}
\label{fig:dist_5-10e10}
\end{center}
\end{figure}

\begin{figure}[t!]
\begin{center}
\includegraphics[width=6cm]{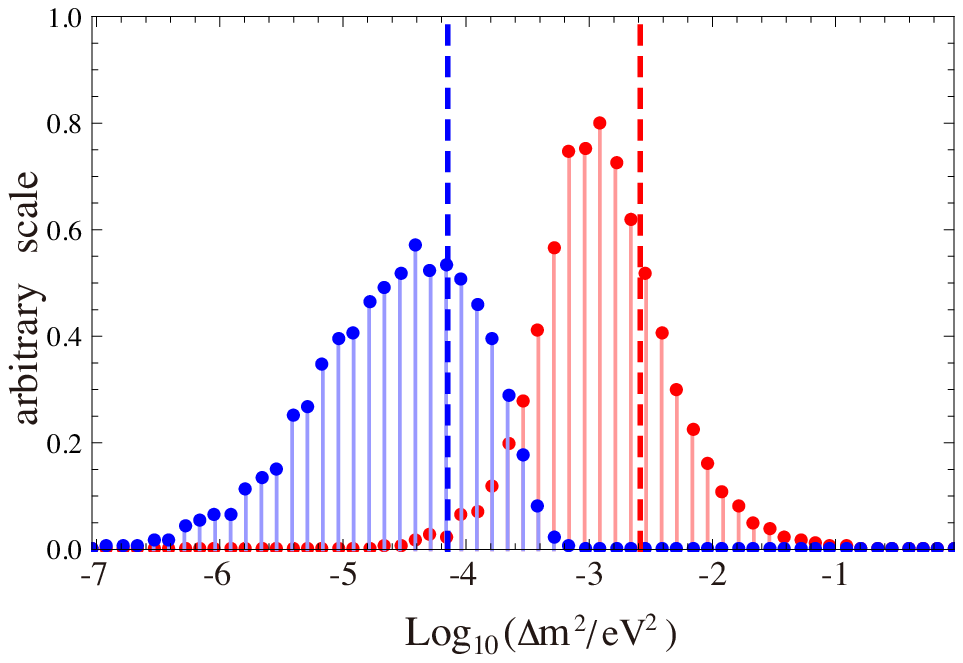}\qquad
\includegraphics[width=6cm]{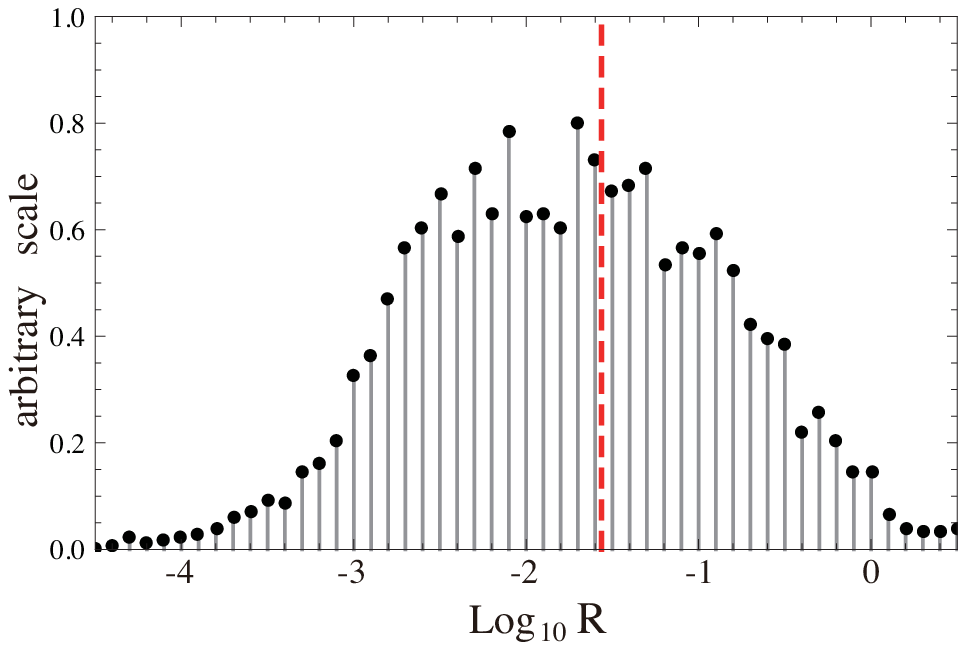}
\vskip 0.5cm
\includegraphics[width=6cm]{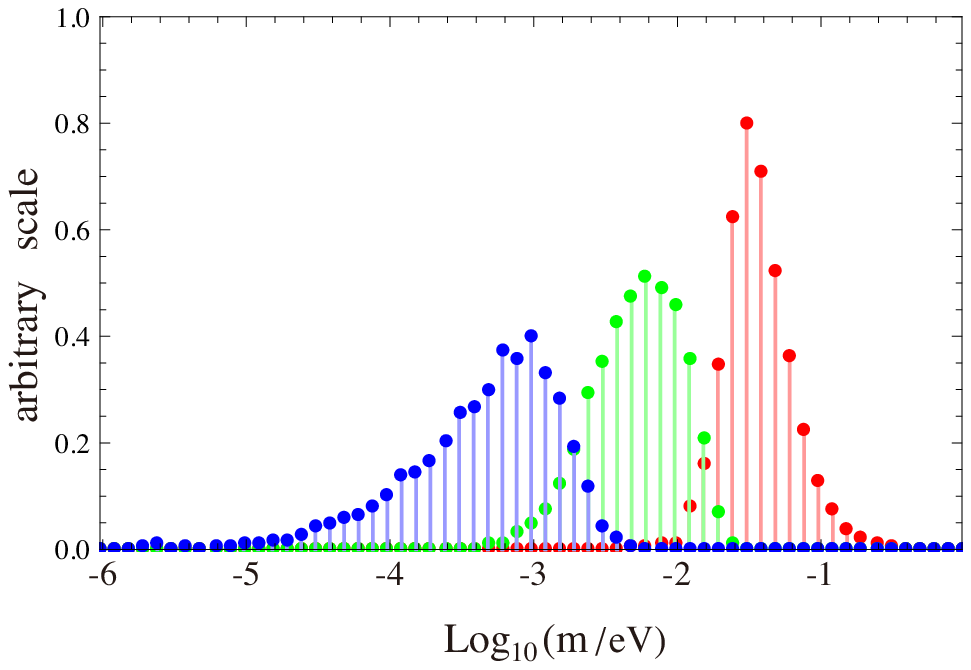}\qquad
\includegraphics[width=6cm]{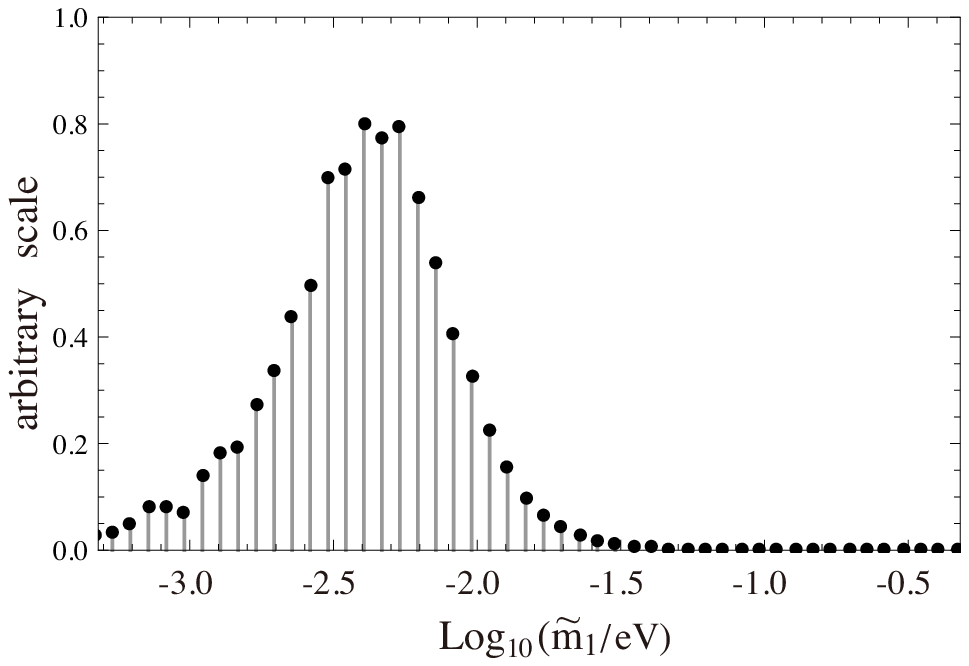}
\caption{
Same as Fig.~\ref{fig:dist_1e15}, except that we have imposed a constraint
$M_1 \leq 10^{10}$\,GeV.
}
\label{fig:dist_10e10}
\end{center}
\end{figure}

Now let us consider a case of $T_R \ll M_0$. In this case, there is an
upper bound on $M_1$ for $N_1$ to be thermally produced. See
\EQ{s1}. To simplify our analysis, we introduce a cut-off on $M_1$,
$M_1 \leq M_{1, max}$, where the maximum value $M_{1, max}$ is
comparable to $T_R$.\footnote{ This is an approximation because the
  coefficient $z$ in \REF{s1} depends on the neutrino Yukawa
  couplings.}

  We show in Fig.~\ref{fig:dist_1e13} the distributions
of the neutrino parameters for $ M_{1, max}= \GEV{13}$. We can see that
the upper bound on ${\widetilde m}_1$ is saturated at about
$O(0.1)$\,eV, which results in the relatively heavy $m_3 \gtrsim 0.1$\,eV. As a
result, the distribution of $\Delta m_{32}^2$ is peaked at $\sim 0.1 {\rm\,eV}^2$
in strong tension with the observations. Also, the distribution of $R$ is peaked below $\sim
10^{-2}$ in slight tension with the observations.
We emphasize here that this tension cannot be removed by simply changing the typical
scale of $M_0$.  If we increase $M_0$, the distribution of $\Delta
m_{21}^2$ decreases almost in proportion to $1/M_0$, because the
effect of $m_2$ and $m_1$ on the leptogenesis is mild. On the other
hand, the distribution of $m_3$ and therefore of $\Delta m_{32}^2$ does
not change significantly, because it is determined by the balance between
leptogenesis and random matrices of order unity. Thus, the distribution of $R$ goes toward
even smaller values, and the tension actually gets severer. For instance,
if we take $M_0 = 3\times \GEV{15}$,
the distribution of $\Delta m_{21}^2$ can be in agreement with the observation,
while that of $R$ is peaked at about $10^{-3}$.
 If we decrease $M_0$, on the other hand, $\Delta m_{21}^2$  increases while
 $\Delta m_{32}^2$ remains almost the same.
  As a result, the mass squared difference $\Delta m_{21}^2$ will be much larger than
the observed values. For instance, both $\Delta m_{21}^2$ and $\Delta m_{32}^2$
are about two orders of magnitude larger than the observed ones for $M_0 = 3 \times
\GEV{14}$.

The situation changes when we consider  $M_{1, max} \lesssim
\oten{11}$\,GeV, for which it becomes difficult to generate the right amount of the baryon asymmetry
and the upper bound on ${\widetilde m}_1$ from
the wash-out effect becomes tighter, ${\widetilde m}_1 \lesssim 0.1$\,eV. The distributions of the neutrino parameters are
shown in Fig.~\ref{fig:dist_5-10e10} for $M_{1, max} = 5 \times
10^{10}$\,GeV.  One can see that the distribution of $R$ is peaked at
about $1/30$, and also the mass squared differences agree very well
with the observations. We also show the distribution of $m_{ee}$.
We have confirmed that the situation is similar for $M_{1, max} = 10^{11}$\,GeV.

The results for $M_{1, max} = \GEV{10}$ are shown in Fig.~\ref{fig:dist_10e10}.
One can see that the distributions are consistent with the observations,
although the constraint on ${\widetilde m}_1$ becomes even tighter
and the distribution of $R$ becomes broader.

To summarize, the neutrino mass distribution nicely explains the
observed neutrino mass squared differences for $M_{1, max} \approx
\oten{10-11}$\,GeV, or equivalently, $T_R = \oten{9-10}$\,GeV in terms of
the reheating temperature. On the other hand, the neutrino mass
distribution is in tension with the observations for $T_R$ higher than
$\oten{11}$\,GeV, and in particular the tension is significant for $T_R$ around
$\oten{12-13}$\,GeV.  The solution of $T_R = \oten{9}$\,GeV is
particularly interesting from the point of view of inflation model
building, dark matter, and the recent indications of a $125$\,GeV
SM-like Higgs boson, which will be discussed in Sec.\ref{sec:4}.

\vspace{5mm}

We close this subsection by briefly discussing the distribution of $m_{ee}$ defined by
\beq
m_{ee} \;=\; \sum_{i=1}^{3} \left(U_{MNS} \right)_{e i}^2 m_i,
\label{mee}
\eeq
to which the amplitude for neutrinoless double beta decay
($0\nu\beta\beta$) is proportional.  The best limit for $^{76}$Ge is
 $|m_{ee}| < 0.35$\,eV~\cite{KlapdorKleingrothaus:2000sn}, and the recent constraints
 for $^{136}$Xe are given by KamLAND-Zen~\cite{KamLANDZen:2012aa} and EXO-200~\cite{Auger:2012ar}
 experiments as, $|m_{ee}| \lesssim (0.3 - 0.6)$\,eV
and  $|m_{ee}| \lesssim (0.14 - 0.38)$\,eV at 90\%C.L.,
respectively.

 From the observed values of the
mixing angles (\ref{mixings}), one finds \bea m_{ee} \simeq 0.67 m_1 +
0.30 m_2 e^{i \alpha_{21}} + 0.02 m_3 e^{i \alpha_{31}}, \eea implying
that $m_{ee}$ is very small for the neutrino masses with normal
hierarchy.  Indeed, $m_{ee}$ typically lies in the range of a few
$10^{-3}$ eV in the neutrino mass anarchy, and thus is below the reach
of current experiments.  Here we have used the fact that the
distribution of $U_{MNS}$ is subject to the invariant Haar measure and
is independent of leptogenesis, and that the neutrino mass eigenvalues
obey the conditional distribution where thermal leptogenesis works
successfully and the observed mass squared differences are realized.

\subsection{Non-thermal leptogenesis}
\label{sec:3-3}
It is not known how the reheating proceeds, because the coupling of
the inflaton with the SM particles are poorly constrained. In a class
of inflation models, the inflaton~\cite{Nakayama:2011ri} or waterfall
field~\cite{hep-ph/9406319} is identified with the U(1)$_{\rm B-L}$
Higgs boson, which is naturally coupled to the right-handed neutrinos
to generate a large Majorana mass.  Then the right-handed neutrinos
are produced by the inflaton decay, and non-thermal leptogenesis takes
place if the reheating temperature is lower than
$M_1$~\cite{Asaka:1999jb,Hamaguchi:2001gw}. The right amount of the
baryon asymmetry can be created at a low reheating temperature, $T_R
\gtrsim \GEV{6}$, since the wash-out effect is suppressed.

Let us consider non-thermal leptogenesis with the neutrino mass
anarchy hypothesis. Suppose that the inflaton mass $m_\phi$ is much
smaller than $M_0$. Then, the typical mass spectrum will be
\beq
T_R < M_1 \lesssim m_\phi \ll  \ M_{2} \leq M_{3} .
\eeq
However, since the wash-out effect is weak at such low $T_R$, the
constraint on the neutrino Yukawa coupling is much weaker than thermal
leptogenesis. Therefore, the contribution of $N_1$ to the light
neutrino mass will be significantly larger than those of $N_2$ and
$N_3$, and the resultant mass spectrum is
\beq
m_1 \leq m_2 \ll m_3,
\eeq
leading to an unacceptably small value of $R$, in contradiction with the
observations.

This problem can be avoided if the inflaton mass is heavier than or
comparable to $M_0$:
\beq
m_\phi \;\gtrsim M_0.
\eeq
For the reference value of $M_0 = \GEV{15}$, this inequality is met
only for a limited class of inflation models such as a smooth hybrid
inflation~\cite{Lazarides:1995vr}.  Note that this problem can be
avoided in a flavor model in which the lightest right-handed neutrino
is charged under a U(1) flavor symmetry.

\section{Discussion and conclusions}
\label{sec:4}

We have so far fixed $M_0 = \GEV{15}$. Let us briefly discuss what
happens if other values of $M_0$ are chosen. First note that normal
mass hierarchy is preferred compared to inverted hierarchy or
degenerate spectrum in the neutrino mass anarchy. Thus, the mass
squared differences $(\Delta m^2_{32}, \Delta m^2_{21})$ are
approximately given by $(m_3^2, m_2^2)$ for most of the cases, and
these two mass scales must agree well with the observations.  For $M_0$
many orders of magnitude smaller than $\GEV{15}$, it is therefore necessary to change the
typical value of the neutrino Yukawa couplings in order to explain the
observed neutrino mass squared differences.  In particular, this is
necessary because otherwise $m_2$ (and therefore $\Delta m^2_{21}$) would be in
tension with the observations.  For example, if $M_0$ is suppressed by
a factor of $10^2$, we should impose ${\rm Tr}[h h^\dag] \leq 10^{-2}$
instead of ${\rm Tr}[h h^\dag] \leq 1$. On the other hand, $M_0$ cannot be much larger than
$\GEV{15}$ because the neutrino Yukawa couplings are required to be
too large for perturbative computations. In these cases it is natural to introduce
a common flavor charge of the right-handed neutrinos to explain the typical values
of $M_0$ and $h_{i \alpha}$. This is an interesting possibility, but it requires computationally
expensive parameter scan. So, here let us focus on the neutrino Yukawa couplings of order
unity and briefly mention how the distribution changes for a slightly different value of $M_0$.
We have studied different values of $M_0$ around $\GEV{15}$, and found that the
allowed region of $M_{1, max}$ decreases as  $M_0$ deviates from $\GEV{15}$.
For instance, the distributions are in agreement with the observations for
$M_{1, max} \approx 3 \times \GEV{11} - \GEV{12}$, if we take $M_0 = 3 \times \GEV{15}$.
As we increase $M_0$ further, $\Delta m_{21}^2$ tends to be too small to account for the
observed value. On the other hand, if we take $M_0 = 3 \times \GEV{14}$, $\Delta m_{21}^2$
tends to be large, and $M_{1, max} \approx \GEV{10}$ gives
distributions barely consistent with the observations. Thus, for $M_0 = 3 \times \GEV{14} \sim 3 \times \GEV{15}$,
the neutrino mass distributions agree well with the observations for $ M_{1, max} = \GEV{10} \sim \GEV{12}$,
or equivalently, $T_R = \oten{9-11}$\,GeV.

The SM with three right-handed neutrinos has been considered in our
analysis, but it is straightforward to extend it to the supersymmetric
(SUSY) framework. Our main conclusion still holds in this case.
Interestingly, the ATLAS and CMS collaborations have recently provided
hints for the existence of a SM-like Higgs particle with mass about
$125$\,GeV~\cite{Higgs-LHC}.  The relatively light Higgs boson mass
suggests the presence of new physics at scales below the Planck
scale~\cite{EliasMiro:2011aa}.  In SUSY extensions of the SM, a 125
GeV Higgs mass can be explained without invoking large stop mixing if
the typical sparticle mass is at $O(10)$\,TeV or heavier.  Among
various possibilities of the SUSY breaking mediation mechanisms, the
simplest one is the anomaly mediation with a generic K\"ahler
potential~\cite{Giudice:1998xp,Ibe:2011aa}.  Then the Wino is likely
the lightest SUSY particle (LSP), and therefore a candidate for dark
matter if the R-parity is conserved. For the gravitino mass of
$O(100)$\,TeV, the thermal relic abundance of the Wino is too small to
account for the observed dark matter density. The correct abundance
can be naturally realized by the gravitino decay, if $T_R =
\oten{9}$\,GeV. Interestingly enough, with this reheating
temperature, the thermal leptogenesis is possible. Furthermore, we
have seen in the previous section that the neutrino mass spectrum is
typical in the conditional distribution at $T_R = \oten{9-10}$\,GeV for
$M_0 \approx \GEV{15}$ when the neutrino mass anarchy is assumed.
Thus, the 125\,GeV SM-like
Higgs boson, the LSP dark matter produced by the gravitino decay,
thermal leptogenesis, and the neutrino mass anarchy point to $T_R =
{\oten{9-10}}$\,GeV.  Such a coincidence is interesting and even
suggestive.

We have assumed that there is no flavor symmetry which distinguishes
three generations of the neutrinos.  The observed hierarchical
spectrum of quarks and charge leptons, on the other hand, can be
nicely explained by the flavor symmetry under which only ${\bf
  10}$-plets are charged while ${\bf 5}^*$-plets are neutral in the
language of the SU(5)~\cite{Hall:1999sn}. Irrespective of whether the
flavor symmetry is a true symmetry or an emergent one, this is
consistent with the SU(5) GUT. We also note that some of the problems
outlined in the previous section (e.g. the difficulty in the
non-thermal leptogenesis) can be easily solved if there is a flavor
symmetry under which the right-handed neutrinos are charged. It should
be emphasized that such flavor symmetry does not affect the see-saw
formula for the light neutrino mass. In fact, it was shown that the
see-saw mechanism is robust against splitting the right-handed
neutrino masses in this way~\cite{Kusenko:2010ik}.

One of the important assumptions in our analysis is that both $h_{i
  \alpha}$ and $X_{ij}$ obey random distribution of order unity. The
typical value can be different from order unity by assigning a common
flavor charge on three generations of $N_i$ and/or $\ell_\alpha$, or
by an extra dimensional set-up. In this sense the neutrino mass
anarchy and the conventional flavor symmetry are compatible.  See
Ref.~\cite{Buchmuller:2011tm} for the recent study of the neutrino
mass anarchy with a certain flavor symmetry~\cite{Buchmuller:1998zf}.

In our analysis we have assumed the linear measure of $h_{i\alpha}$ and $X_{ij}$.
It is in principle possible to adopt another measure which depends on a U(3)-invariant
factor, such as ${\rm Tr}[h h^\dag]$ or ${\rm det}[h]$. In this case, while the
mixing angles and CP violation phases are still determined by the invariant
Haar measure, the distributions of the neutrino masses are generically modified.
However, we believe that our results are robust against changing
the measure to some extent. This is because the peak position of  $m_3$ is determined
by the balance between the randomness and the wash-out effect. 
To see this, let us consider thermal leptogenesis with a reheating temperature lower than
the typical right-handed neutrino mass.  Recall that  the successful thermal 
leptogenesis selects the subset
${\cal S}_1 \bigcap {\cal S}_2$, where  $M_1 \ll M_{2}, M_3$ and
$|h_{1 \alpha}| \ll 1$ are realized. Therefore, for given $T_R$,
the peak of the distribution of $m_3$ is not sensitive to the details of the adopted measure,
as long as the conditional distribution of $M_1$ is peaked near $T_R$ and the wash-out bound
on the $|h_{1 \alpha}|$ is saturated. Suppose, for instance, that 
the measure is an increasing function of the eigenvalues of $h$ and $M$; then
the upper bounds on $M_1$ and $h_{1 \alpha}$ are considered to be saturated
because it requires severe fine-tunings to realize  smaller values.
Therefore, as long as the measure of $h_{i\alpha}$ and $X_{ij}$ satisfy such property,
the distribution of $m_3$ is not modified significantly. The distribution of $m_1$ and
$m_2$ are not sensitive to the leptogenesis, and so, our results are valid if
the measure  is such that the predicted neutrino mass spectrum agrees with the observation when
successful leptogenesis is not imposed. Of course, our results do not hold, for instance,
if the measure favors a lighter right-handed neutrinos or smaller neutrino Yukawa couplings.

The observed neutrino mixing angles and mass squared difference
support both the neutrino mass anarchy and the see-saw mechanism.
This suggests that the neutrino masses and mixings are irrelevant to the existence
of life and therefore keeps its original distribution in the
landscape.  On the other hand, however, if the leptogenesis is
responsible for the origin of matter, it selects a certain subset of
the neutrino parameters, which may significantly distort the original
distribution spoiling the success of neutrino mass anarchy.  In this
paper we have studied if the successful leptogenesis is possible together with the neutrino mass anarchy
hypothesis.  We have found that the distributions of the neutrino mixing angles as well as Dirac and
Majorana CP violation phases are determined by the invariant Haar
measure of U(3), even if we impose successful leptogenesis. On the other hand,
the neutrino mass spectrum is generically affected by
leptogenesis. In the case of thermal leptogenesis, the mass spectrum
for the right-handed neutrinos and the neutrino Yukawa matrix exhibit
a certain pattern, as a result of the competition between random
matrices with elements of order unity and the wash-out effect. The
hierarchical mass spectrum for the right-handed neutrinos as well as
suppressed neutrino Yukawa couplings $h_{1 \alpha}$ are similar to
that obtained by an approximate flavor symmetry. In a sense, the
flavor symmetry is emergent.  However, as we look into details, we
have seen that the neutrino mass spectrum depends sensitively on the
reheating temperature.  The light neutrino mass distribution is
consistent with observation if the reheating temperature is
$\oten{9-11}$\,GeV for $M_0 = 3\times \GEV{14} \sim 3\times \GEV{15}$.
In particular, the solution of $T_R = \oten{9}$\,GeV is interesting in
connection with the neutralino dark matter produced by the gravitino
decay and the 125\,GeV SM-like Higgs boson.  On the other hand,
non-thermal leptogenesis is consistent with the observations only if
the inflaton mass is heavier than the typical right-handed neutrino
mass scale.  Such a heavy inflaton mass can be realized in e.g. the
smooth hybrid inflation.

\section*{Acknowledgment}
FT thanks N. Haba and K. Nakayama for communication.
 This work was supported by the Grant-in-Aid for Scientific Research
 on Innovative Areas (No.24111702, No.21111006 and No.23104008) [FT],
 Scientific Research (A) (No.22244030 and No.21244033 [FT]), and JSPS
 Grant-in-Aid for Young Scientists (B) (No.24740135) [FT].  This work
 was also supported by World Premier International Center Initiative
 (WPI Program), MEXT, Japan, and by Grant-in-Aid for Scientific
 Research from MEXT (No. 23104008 and No. 23540283 [KSJ]).

\end{document}